\documentclass[journal]{IEEEtran}
\usepackage{cite}
\usepackage{amssymb}
\usepackage{amsfonts}
\usepackage{fancyhdr}  %
\usepackage{extarrows}
\usepackage{algorithm}
\usepackage{multirow,tabularx}
\usepackage{mathtools}
\usepackage{xcolor}
\usepackage{bm}
\usepackage{algorithmic}
\usepackage{amsmath}
\usepackage{subfigure}
\usepackage{makecell}
\usepackage{colortbl}
\usepackage{bbm}
\usepackage{enumitem}

\allowdisplaybreaks

\makeatletter
\let\markboth\@gobbletwo
\makeatother
\usepackage{titlesec}
\renewcommand{\thesubsubsection}{\arabic{subsubsection}}

\titlespacing{\section}{0pt}{1ex}{0.5ex}
\titlespacing{\subsection}{0pt}{1ex}{0.5ex}

\titleformat{\subsubsection}[runin]
{\normalfont\itshape}
{\thesubsubsection)}{0.5em}{}[: \hspace{0.5em}]

\ifCLASSINFOpdf
  \usepackage[pdftex]{graphicx}
  \graphicspath{{../pdf/}{../jpeg/}}
  \DeclareGraphicsExtensions{.pdf,.jpeg,.png}
\else
  \usepackage[dvips]{graphicx}
  \graphicspath{{../eps/}}
  \DeclareGraphicsExtensions{.eps}
\fi
\makeatletter
\newenvironment{breakablealgorithm}
  {% \begin{breakablealgorithm}
   \begin{center}
     \refstepcounter{algorithm}% New algorithm
     \hrule height.8pt depth0pt \kern2pt% \@fs@pre for \@fs@ruled
     \renewcommand{\caption}[2][\relax]{% Make a new \caption
       {\raggedright\textbf{\ALG@name~\thealgorithm} ##2\par}%
       \ifx\relax##1\relax % #1 is \relax
         \addcontentsline{loa}{algorithm}{\protect\numberline{\thealgorithm}##2}%
       \else % #1 is not \relax
         \addcontentsline{loa}{algorithm}{\protect\numberline{\thealgorithm}##1}%
       \fi
       \kern2pt\hrule\kern2pt
     }
  }{% \end{breakablealgorithm}
     \kern2pt\hrule\relax% \@fs@post for \@fs@ruled
   \end{center}
  }
\makeatother

\begin{document}

\title{NLOS-Aided Joint OTA Synchronization and Off-Grid Imaging for Distributed MIMO Systems}

\author{Xin~Tong,~\IEEEmembership{Member,~IEEE, }
        Lechen~Zhang,~\IEEEmembership{Graduate Student Member,~IEEE, }
        Yu~Ge,~\IEEEmembership{Member,~IEEE, }
        Dario~Tagliaferri,~\IEEEmembership{Member,~IEEE, }
        Henk~Wymeersch,~\IEEEmembership{Fellow,~IEEE } 
        % <-this % stops a space

\thanks{Xin~Tong, Yu~Ge and Henk~Wymeersch are with the Department of Electrical Engineering, Chalmers University of Technology, Gothenburg, Sweden. (emails: \{xinto, yuge, henkw\}@chalmers.se).}

\thanks{Lechen~Zhang is with the National Key Laboratory of Wireless Communications, University of Electronic Science and Technology of China, Chengdu, China (e-mail: zhanglechen@std.uestc.edu.cn).}

\thanks{Dario~Tagliaferri is with the Department of Electronics, Information and Bioengineering, Politecnico di Milano, 20133, Milano, Italy (e-mail: dario.tagliaferri@polimi.it).}

\thanks{This work was supported, in part, by the SNS JU project 6G-DISAC under the EU’s Horizon Europe research and innovation Program under Grant Agreement No 101139130 and the Swedish Research Council (Grant 2022-03007).}

% \thanks{This work was supported in part by National Natural Science Foundation of China under Grants 62394292 and U20A20158, Zhejiang Provincial Key R\&D Program under Grant 2023C01021, Ministry of Industry and Information Technology under Grant TC220H07E, and the Fundamental Research Funds for the Central Universities under Grant No. 226-2024-00069, and the SNS JU project 6G-DISAC under the EU’s Horizon Europe research and innovation Program under Grant Agreement No 101139130.}
}

\maketitle

\begin{abstract}
Distributed multiple-input multiple-output (MIMO) architectures enable large-scale integrated sensing and communication (ISAC) by providing high spatial resolution and robustness through spatial diversity. However, practical phase-coherent sensing is challenged by phase synchronization errors and modeling mismatch caused by grid discretization. Existing over-the-air (OTA) synchronization methods typically treat synchronization and sensing tasks separately, which may lead to inaccurate phase alignment when multipath components are used for imaging.
In this paper, we propose a non-line-of-sight (NLOS)-aided joint OTA synchronization and off-grid imaging framework for distributed MIMO ISAC systems. First, a line-of-sight (LOS)-assisted coarse synchronization is performed to establish initial phase coherence across distributed links. Subsequently, an iterative refinement stage exploits reconstructed NLOS components obtained from imaging results. 
By modeling off-grid effects via a first-order Taylor expansion, we transform measurements with nonlinear off-grid offset into an augmented linear model with jointly sparse reflectivity and off-set variables.
The imaging problem is reformulated as a structured sparse recovery task and solved using a tailored off-grid approximate message passing (OG-AMP) algorithm. The imaging and synchronization modules are coupled within a closed-loop alternative optimization framework, where improved imaging enables more accurate phase refinement, and vice versa.
Numerical results show that the proposed framework achieves accurate synchronization and imaging under phase errors. Compared with conventional approaches, it shows superior robustness and accuracy.
\end{abstract}

\begin{IEEEkeywords}
  Distributed multiple-input multiple-output (MIMO), integrated sensing and communication (ISAC), over-the-air (OTA) synchronization, off-grid imaging.
\end{IEEEkeywords}

\IEEEpeerreviewmaketitle

\section{Introduction}
%\subsection{Motivation}
\IEEEPARstart{I}{n} future integrated sensing and communication (ISAC) systems, the distributed multiple-input multiple-output (MIMO) architecture is considered as a key factor in achieving large-scale environment sensing, supporting applications such as intelligent transportation, extended reality, and ubiquitous environmental monitoring \cite{Liu, Behdad}. In such systems, large-scale deployment of spatially separated transmitters (Txs) and receivers (Rxs) enables unprecedented sensing resolution and communication reliability. In practice, such systems operate in complex propagation environments where line-of-sight (LOS) and abundant non-line-of-sight (NLOS) multipath components coexist. While this poses challenges for over-the-air (OTA) synchronization and phase-coherent sensing, the joint presence of LOS and NLOS paths also provides new opportunities to exploit multipath diversity for synchronization and sensing within a unified OTA framework \cite{han2025over, yibo}.

The first challenge arises from the tight synchronization requirement in distributed MIMO systems \cite{Emil}. Each transceiver has its own clock and oscillator, leading to time offsets (TOs) and phase offsets (POs), which can have an impact on multiple links. These residual offsets distort phase-sensitive measurements and severely degrade coherent processing such as beam-focusing or multi-static imaging \cite{Aguilar}. 
Although wired synchronization or centralized clock distribution may appear as potential solutions, their achievable accuracy is generally insufficient for large-scale phase-coherent sensing, and their deployment is impractical in flexible and distributed ISAC architectures. As a result, synchronization should be achieved in an OTA framework based on wireless measurements. 
In practice, LOS paths are commonly exploited to provide a coarse synchronization reference. However, LOS-based synchronization alone is often unreliable in realistic distributed MIMO scenarios due to occlusions, intermittent visibility, and limited geometric diversity \cite{Mike, Larsson, Pegoraro}. Thus, there is a strong need for robust synchronization that can operate with abundant NLOS links.

Another challenge arises from POs introduced by the mismatch between grid-based imaging models and the continuous physical environment. In large-scale sensing, the region of interest is commonly discretized into coarse grid of pixels to control computational complexity \cite{tongxicc}. As a result, the main scattering centers rarely coincide with grid center points, leading to off-grid modeling errors. For imaging, such off-grid offsets partly manifest as phase variations in the scattering coefficients and also affect the accuracy of coherent processing. Moreover, when the reconstructed data is further used for refined synchronization, these off-grid-induced POs are fundamentally indistinguishable from those caused by imperfect synchronization, making coherent processing particularly challenging \cite{Dai}.
While using extremely fine grids could reduce such phase errors, the resulting memory and computational overhead makes this method impractical. Importantly, in NLOS-rich environments, multipath propagation will provide more geometrically diverse measurements that can be used for joint synchronization and imaging.

Motivated by these challenges, this paper proposes a NLOS-aided ISAC framework that integrates LOS-based coarse synchronization with NLOS-assisted joint refinement and off-grid imaging in distributed MIMO systems.  The proposed approach leverages multipath for synchronization, achieves high-resolution off-grid imaging, and is suitable for practical ISAC deployments.

\subsection{Related Works}
From the perspective of sensing methods in ISAC systems, conventional imaging techniques have been widely studied, such as radar-based imaging \cite{liuf2, Bi}, phased-array beamspace imaging \cite{che}, and multistatic coherent imaging \cite{Manzoni, Tagliaferri}. Compared with pure localization or parameter estimation, imaging-based sensing provides a richer and more interpretable representation of the environment, enabling the extraction of spatial structures, target shapes and sizes, and scattering distributions that are highly beneficial for ISAC. However, these model-driven linear imaging methods are limited by the sidelobe structure of the resulting point spread function, while computational imaging methods can improve performance by leveraging sparsity priors \cite{liua}. Although distributed ISAC architectures provide increased spatial degrees of freedom through geographically separated nodes, classical linear imaging methods do not fully exploit this advantage. In dense NLOS environments commonly encountered in ISAC systems, classical imaging models either become inaccurate or require increasingly complex propagation modeling, and the resulting images often lack fine spatial resolution.

To overcome these limitations, computational imaging has recently emerged as a powerful alternative for wireless environment sensing \cite{Sun}. By dividing the scenario into discrete pixels and recovering their scattering coefficients, computational imaging transforms the sensing task into a sparse reconstruction problem that can achieve high-resolution imaging even with limited physical resources. In addition, compared to classical linear imaging methods, computational imaging provides a more flexible framework and accurate results for ISAC. By passively reusing communication signals, it avoids interference with the communication process. Based on compressed sensing theory \cite{Candes}, various sparse reconstruction algorithms, such as the orthogonal matching pursuit (OMP) algorithm \cite{omp} and the generalized approximate message passing (GAMP) algorithm \cite{Rangan}, have been applied to MIMO orthogonal frequency division multiplexing (OFDM) sensing \cite{linz} and millimeter-wave communication environment reconstruction \cite{tongx, tongx2}. These methods demonstrate the great potential of computational imaging as a fundamental feature of future ISAC systems.

Despite these advantages, existing computational imaging methods are typically developed under the assumption of ideal synchronization. In practical distributed ISAC systems, TOs and POs can severely degrade imaging performance. To address this issue, map-assisted or multipath-assisted synchronization techniques have been proposed to improve coherence among distributed transceivers. For instance, \cite{han2025over} shows that environmental information can effectively reduce time and phase misalignment in distributed MIMO systems, with related ideas applied to cell-free massive MIMO \cite{Emil}, multistatic radar \cite{Bryant}, and cooperative localization \cite{ojas}. However, existing synchronization approaches are generally developed independently of imaging and often rely on either strong LOS components or accurate prior map information. The lack of a unified framework that jointly integrates imaging and synchronization limits their applicability in practical large-scale ISAC systems.

\subsection{Contributions}
In this paper, we propose joint OTA synchronization and environment imaging in distributed MIMO ISAC systems operating in NLOS-rich scenarios. Different from conventional approaches that address synchronization and sensing separately, we aim to exploit abundant NLOS multipath components as a shared resource for both synchronization and imaging.
Specifically, we first formulate a unified OFDM signal model that captures both synchronization POs and off-grid scattering effects within a common computational imaging framework. We exploit available LOS components to obtain a coarse estimation of POs across distributed transceivers. Building upon this initial synchronization, we perform sparse environment imaging using NLOS multipath components. Based on message passing theory, off-grid approximate message passing (OG-AMP) algorithm is proposed to reconstruct sparse NLOS scattering structures. This imaging result enables the estimation and refinement of OTA synchronization parameters. The updated synchronization parameters in turn improves phase coherence across links, leading to progressively more accurate imaging results. Through this alternative optimization (AO) between imaging and synchronization, the proposed method enables reliable joint sensing and synchronization in ISAC systems. 
The main contributions of this work are summarized as follows:

\begin{itemize}
    \item We propose a unified OFDM-based ISAC framework that jointly performs OTA phase synchronization and environment imaging by passively exploiting channel state information (CSI) obtained from the standard communication process. Instead of treating NLOS multipath as interference, the proposed method leverages those paths as virtual anchors to support both synchronization and sensing in distributed MIMO systems.

    \item An OG-AMP imaging algorithm is proposed to explicitly account for model mismatch caused by coarse grids and residual synchronization errors. The proposed method is able to reliably recover sparse scattering structures even when phase-level errors are present, which is particularly important in large-scale NLOS scenarios.

    \item We further propose an OTA synchronization strategy that constructs an ideal reference channel from the reconstructed NLOS imaging results. By comparing the received signals with this reference, residual POs can be iteratively refined, forming an AO between imaging and synchronization.

    \item Simulation results demonstrate that, in scenarios where conventional processing methods fail or provide unreliable estimates, the proposed approach achieves effective performance gains for both synchronization and imaging, highlighting its practical value for ISAC deployments.
\end{itemize}

% The rest of this paper is organized as follows. Section \uppercase\expandafter{\romannumeral2} presents the environment setting and system model in the large-scale communication scenario. Section \uppercase\expandafter{\romannumeral3} proposes the multi-view environment sensing algorithm. In Section \uppercase\expandafter{\romannumeral4}, we analyze the impact of the proposed pixel division method and multi-view sensing method on the system performance. Finally, Section \uppercase\expandafter{\romannumeral5} presents the numerical results, and Section \uppercase\expandafter{\romannumeral6} concludes the paper. 

% \textit{Notation}: Fonts $a$, $\bm a$, and $\mathbf{A}$ represent scalars, vectors, and matrices, respectively. $\mathbf{A}^{\mathsf{T}}$ and $\|\mathbf{A}\|_{\rm F}$ denote transpose and Frobenius norm of $ \mathbf{A} $, respectively. $|\cdot|$ and $[\cdot]$ denote the modulus and the catenation of the matrix, respectively. $\odot $ represents the Hadamard product between two matrices. Finally, notation ${\rm diag}({\bm a})$ represents a diagonal matrix with the entries of ${\bm a}$ on its main diagonal, and $\delta(\cdot)$ is the Dirac delta function. 

\section{System Model}
\subsection{Environment Setting}
\begin{figure}[t]
    \centering
    \includegraphics[width=0.7\linewidth]{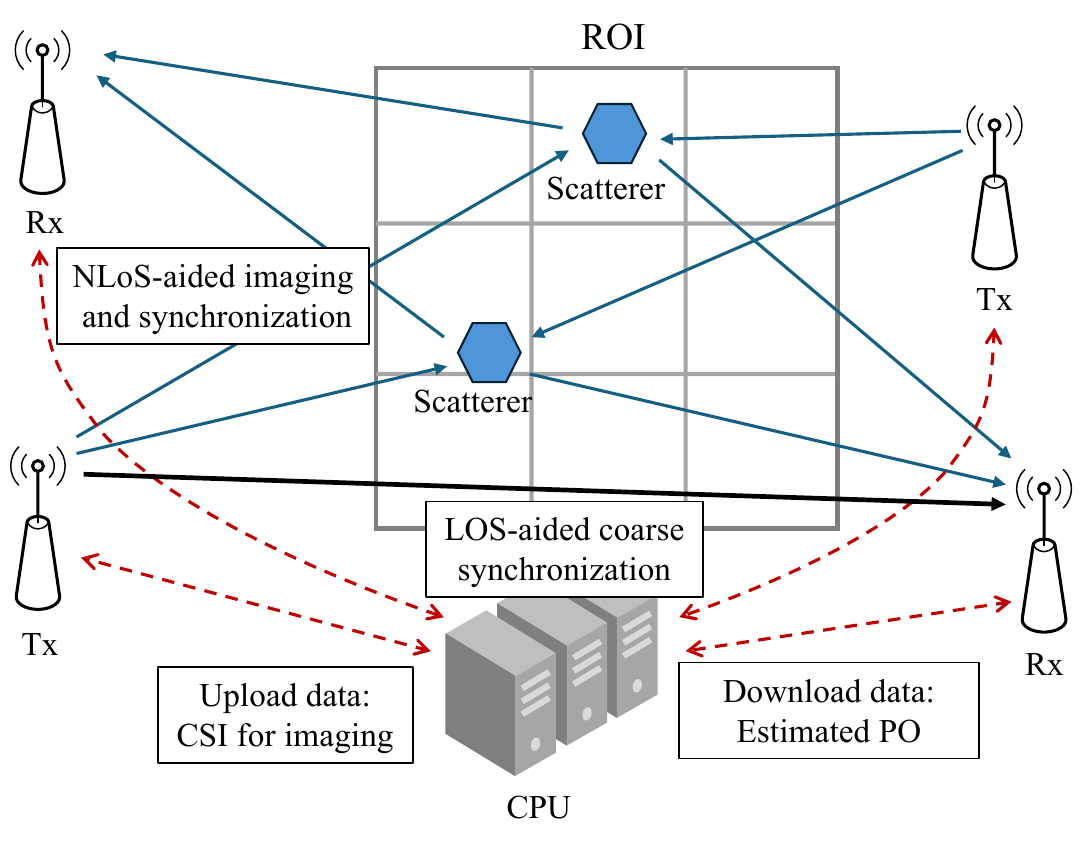}
    \caption{A joint OTA synchronization and imaging scenario in disributed MIMO systems.}
    \label{fig:scenario}
\end{figure}

We consider the distributed MIMO ISAC system in Fig.~\ref{fig:scenario}, including multiple spatially distributed Tx and Rx nodes, each with a single, isotropic antenna. All antennas are connected to a central processing unit (CPU). All Tx and Rx nodes are connected through the available links and the CPU acts as a fusion center, aggregating the CSI from all distributed links to perform joint high-resolution imaging. The imaging results are then assist with OTA synchronization and the estimated synchronization errors are distributed to Txs and Rxs.

In the distributed architecture, each ISAC node operates with an independent local oscillator (LO). The independent LOs in distributed nodes introduce relative clock offsets due to hardware imperfections. As a result, each Tx-Rx link presents an unknown TO, which further induces a random uniformly-distributed carrier PO after downconversion. These POs are absorbed into the effective response and severely degrade the phase coherence required for high-resolution imaging.

In conventional OFDM communication systems, synchronization is typically performed on each Tx-Rx link basis. Coarse timing estimation is achieved using cyclic prefixes (CPs) or predefined preambles, followed by frequency-domain channel estimation and data demodulation, which absorbs both LOS and NLOS multipath components into composite subcarrier coefficients \cite{Beek}. This level of synchronization is generally sufficient for communication systems, where knowledge of absolute delays and phases is not required, as long as these impairments remain quasi-static and channel reciprocity holds.
In contrast, distributed MIMO sensing requires synchronization across multiple Txs and Rxs. Coherent imaging critically relies on consistent phase information across links and subcarriers. 

\subsection{Grid-Based Signal Model}\label{sec:giridmodel}

Each Tx employs a standard OFDM signaling scheme, and transmits data symbols together with predefined pilot symbols over $K$ orthogonal subcarriers with subcarrier spacing $\Delta f$ and a CP. 
Different Txs are multiplexed using orthogonal communication resources so that their individual CSI can be reliably estimated at the Rxs following conventional OFDM-based channel estimation procedures. No sensing-oriented waveform or dedicated probing signal is designed in this work, focusing on a complete reuse of OFDM signals. The proposed framework passively exploits the CSI obtained from the standard communication process. 

As shown in Fig.~\ref{fig:scenario}, we discretize the 2-dimensional (2D) region of interest (ROI) into pixels, where each pixel represents the mean centroid of scatterers and an average scattering coefficient within that pixel.\footnote{The analysis method for 3-dimensional (3D) scenarios can be derived using a similar approach.}
We use the scattering coefficient $x_n$ to represent the environmental information in the $n$-th pixel.
If no target object in $n$-th pixel, $x_n = 0$. Otherwise, we have $x_n \in (0,1]$, which is normalized according to a maximum value of expected environment reflectivity and the total number of pixels is denoted as $N$.
Therefore, the environmental information of the ROI can be characterized by the scattering coefficient vector ${\bm{x}} = [x_1, x_2, \ldots, x_{N}]^{\mathsf T}$. It is worth noting that ${\bm{x}}$ is a property of the environment itself, not related to the propagation path.

Let the center of the $n$-th pixel be $\bm p_n=[x_n,y_n]^{\mathsf T}$, Tx position $\bm p_{{\rm Tx},m}=[x_{{\rm Tx},m},y_{{\rm Tx},m}]^{\mathsf T}$ and Rx position $\bm p_{{\rm Rx},m}=[x_{{\rm Rx},m},y_{{\rm Rx}.m}]^{\mathsf T}$. The free-space propagation channel gain $a_{m, k}(\bm p_n)$ passing through the $n$-th pixel of the $m$-th Tx-Rx link is calculated as \cite{goldsmith2005wireless}
\begin{equation}\small
    a_{m, k}(\bm p_n) = \frac{1}{L_m(\bm p_n)}\exp(-j 2 \pi f_{k}\tau_m(\bm p_n)),
    \label{eq:fr}
\end{equation}
where $f_{k} = f_c + k \Delta f$ is the frequency of the $k$-th subcarrier with $f_c$ denotes carrier frequency. The propagation delay $\tau_m(\bm p_n)$ and the amplitude attenuation $L_m(\bm p_n)$ are 
{\small\begin{align}
\tau_m(\bm p_n) &= ({\|\bm p_n - \bm p_{{\rm Tx},m} \|_2 + \|\bm p_n - \bm p_{{\rm Rx},m} \|_2})/{c}, \\
L_m(\bm p_n) &= 4\pi\alpha({\|\bm p_n - \bm p_{{\rm Tx},m} \|_2 \cdot \|\bm p_n - \bm p_{{\rm Rx},m} \|_2})/\lambda_{k},
\end{align}}\noindent
which is assumed to be constant within the bandwidth, $\lambda_{k}$ is the wavelength and $\alpha$ is the normalization constant that implicitly apply to $\bm x$.

Synchronization offsets originate at the node level rather than at the individual Tx-Rx link level. Specifically, we define the node-level baseband POs on the $k$-th subcarrier as \cite{yibo}
{\small \begin{align}
\phi_{{\rm Tx},i,k} &= 2\pi f_{k}\,\delta t_{{\rm Tx},i} + \phi_{{\rm LO},{\rm Tx},i}, \\
\phi_{{\rm Rx},j,k} &= 2\pi f_{k}\,\delta t_{{\rm Rx},j} + \phi_{{\rm LO},{\rm Rx},j},
\end{align}}\noindent
where $\delta t_{{\rm Tx},i} \sim {\rm U}(0,1/\Delta f) $ and $\delta t_{{\rm Rx},j} \sim {\rm U}(0,1/\Delta f)$ denote the TOs of the $i$-th Tx and $j$-th Rx nodes, respectively, and $\phi_{{\rm LO},{\rm Tx},i} \sim {\rm U}(0,2\pi)$ and $\phi_{{\rm LO},{\rm Rx},j} \sim {\rm U}(0,2\pi)$ represent their LO-induced POs. 

In this paper, we will estimate the global PO at the node level to achieve synchronization. The Tx PO enters the baseband signal through upconversion, while the Rx PO appears with an opposite sign due to downconversion using the conjugate local oscillator. The composite PO associated with the $(i,j)$-th Tx-Rx link can then be expressed as \cite{Unnikrishnan}
\begin{equation}\small
\Phi_{m,k} 
= \phi_{{\rm Tx},i,k} - \phi_{{\rm Rx},j,k},
\label{eq:link_phase_node}
\end{equation}
where $m$ indexes the corresponding Tx-Rx link. We consider phase errors induced by node positioning inaccuracies are absorbed into this term. Considering that the proposed method is a single-snapshot imaging method, residual carrier frequency offset (CFO) is incorporated into the constant PO.
By stacking the node-level phase terms into a vector $\boldsymbol{\phi}_k = [\phi_{{\rm Tx},1,k},\ldots,\phi_{{\rm Tx},N_{\rm Tx},k}, \phi_{{\rm Rx},2,k},\ldots,\phi_{{\rm Rx},N_{\rm Rx},k} ]^{\mathsf T} \in \mathbb{C}^{N_{\rm Tx}+N_{\rm Rx}-1}$, 
the link-level phase vector $\boldsymbol{\varphi}_{k} = [\Phi_{1,k}, \ldots, \Phi_{M,k}] \in \mathbb{C}^{M}$ is written as
\begin{equation}\small
\boldsymbol{\varphi}_{k} = \mathbf{G}\boldsymbol{\phi}_k
\label{eq:association_matrix}
\end{equation}
where $\mathbf{G} \in \{ 0, \pm 1 \}^{M \times (N_{\rm Tx}+N_{\rm Rx}-1)}$ is the network incidence matrix, indicating that there are links (LOS or NLOS) between the corresponding Tx and Rx nodes, $M = N_{\rm Tx} \times N_{\rm Rx}$ is the number of links. For $m$-th link connecting Tx $i$ and Rx $j$, the element corresponding to Tx node $i$ is $+1$ and the element corresponding to Rx node $j$ is $-1$. All other elements in $m$-th row are $0$. It is worth noting that the absolute phase reference is unobservable, and therefore the phase of the node with the most links can be fixed as a reference phase. Consequently, the total number of independent phase parameters is reduced from the number of nodes to $N_{\rm Tx}+N_{\rm Rx}-1$.
The CSI $y_{m,k}$ of the $m$-th Tx-Rx can be modeled as\footnote{Higher-order scattering effects are ignored and treated as noise due to their larger attenuation.}
\begin{equation}\small
    y_{m,k} = \Big(h_{m,k}^{\rm LOS}u_m + \sum_{n=1}^{N} a_{m,k}(\bm p_n)x_n \Big) \cdot e^{- j\Phi_{m, k}} + n_{m,k}, \label{eq:ymnf}
\end{equation}
where $u_m$ is a known boolean visibility factor indicating whether the $m$-th link has a LOS path, $n_{m,k}$ represents the Gaussian noise. The LOS channel $h_{m,k}^{\rm LOS}$ is under free-space propagation assumption, as
\begin{equation}\small
    h^{\rm LOS}_{m, k} = \frac{1}{L^{\rm LOS}_m}\exp(-j 2 \pi f_{k}\tau^{\rm LOS}_m),
    \label{eq:hLOS}
\end{equation}
where $\tau^{\rm LOS}_m = {\|\bm p_{{\rm Tx},m} - \bm p_{{\rm Rx},m} \|_2}/{c}$ is the LOS propagation delay. The LOS amplitude attenuation is calculated as $L^{\rm LOS}_m = 4\pi{\|\bm p_{{\rm Tx},m} - \bm p_{{\rm Rx},m} \|_2}/\lambda_{k}$, as for free-space propagation. The LOS path is not affected by scatterers and is therefore independent of the pixel grid.

In this paper, conventional channel estimation and equalization are performed to reliably recover the communication data \cite{Beek}. The estimated CSI on all subcarriers is then passively reused for imaging, enabling ISAC without introducing additional signaling overhead. We jointly process channel estimation results of $M$ Tx-Rx links and $K$ subcarriers, so \eqref{eq:ymnf} can be written in matrix form as
\begin{equation}\small
    \bm y  = \mathbf{\Phi}(\bm{h}^{\rm LOS} \circ {\tilde{\bm u}} + \mathbf A \bm x) + \bm n,
    \label{eq:md}
\end{equation}
where $\bm y \in \mathbb{C}^{MK}$ is a stacked vector of all estimated channel propagation gains, $\bm{h}^{\rm LOS}\in \mathbb{C}^{MK}$ is the stacked vector of $h_{m,k}^{\rm LOS}$. Vector $\tilde{\bm u} = \bm u \otimes \mathbf 1_K \in \mathbb{B}^{MK}$ and $\bm{u} \in \mathbb{B}^{M}$ is the stacked vector of $u_{m}$. Matrix $\mathbf{\Phi} = {\rm diag}([e^{-j\bm \varphi_{1}}, \dots, e^{-j\bm \varphi_{K}}]) \in \mathbb{C}^{MK \times MK}$ is a diagonal matrix containing the POs on all Tx-Rx links and subcarriers. The free-space propagation gain matrix for all links and subcarriers is denoted as $\mathbf A \in \mathbb{C}^{MK \times N}$, used as the reference channel dictionary for imaging.

In this paper, we focus on the joint performance of OTA synchronization and imaging. As shown in \eqref{eq:md}, we consider the POs $\mathbf \Phi$ and environment information ${\bm x}$ as unknown variables to be solved, and the other reference propagation gains are known values.
In addition, some novel propagation models have been proposed to more accurately describe wireless signal propagation in pixelated environments \cite{tongx, tongx2, tongxicc, xing}. These models can be used to replace the $\mathbf A$ model in \eqref{eq:fr} and extended to an off-grid model. For the propagation model, we focus on isotropic targets and we can apply other accurate models and consider other physical properties of pixels (e.g., shape of target, frequency dependence) in the future.

\begin{figure}[t]
    \centering
    \includegraphics[width=0.6\linewidth]{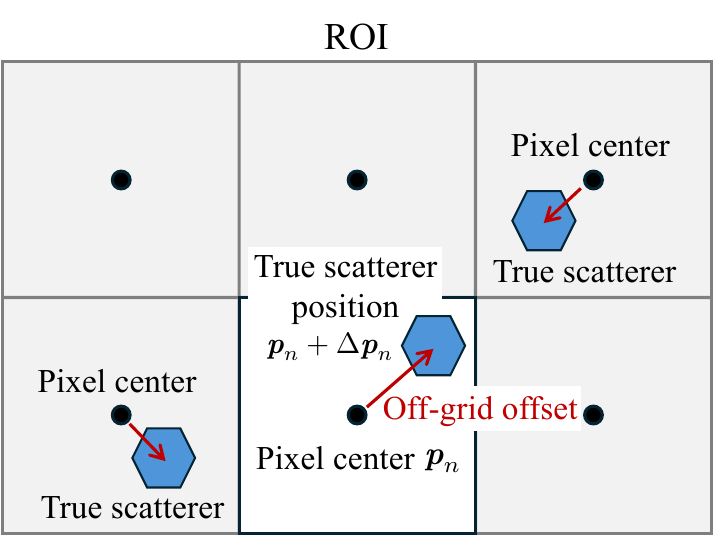}
    \caption{A 2D pixel grid, with targets randomly distributed on the grid, exhibits the off-grid effect shown by the red arrow.}
    \label{fig:og}
\end{figure}

\subsection{Off-Grid Approximation Model}\label{sec:OGC}
In this section, we take a point scatterer as an example to illustrate the off-grid problem in imaging and propose an off-grid channel model as an extension of the grid-based channel model in Section~\ref{sec:giridmodel}. Note that a pixel is the smallest unit of imaging. When there are multiple scatterers in a pixel, these sub-scatterers can be modeled as a single equivalent point scatterer with average intensity and average off-grid offset through coherent superposition of signals \cite{tongxtwc}. As shown in Fig.~\ref{fig:og}, the true scatterer is not exactly at the pixel center, which causes the different PO of each link. Consider a true scatterer located at $\bm p_n + \Delta \bm p_n$ instead of placed in the pixel center $\bm p_n$, where $\Delta \bm p_n$ is the off-grid offset. The reference propagation gain in \eqref{eq:fr} is approximated by first-order Taylor expansion 
\begin{equation}\small
    a_{m, k}(\bm p_n + \Delta \bm p_n) \approx a_{m, k}(\bm p_n) +
    \left( \nabla_{\bm p} a_{m, k}(\bm p_n) \right)^{\mathsf{T}} 
    \Delta \bm p_n.
    \label{eq:Taylor}
\end{equation}
Since the amplitude change is not significant under relatively small off-grid offsets $\Delta \bm p_n$, we consider only variation of the phase information \cite{goldsmith2005wireless}. Inspired by diffraction theory that the spatial gradient is determined by the transmit and receive wavevectors \cite{Manzoni}, the gradient of the position $\bm p_n$ is approximated as
{\small \begin{align}
    \nabla_{\bm p} &a_{m, k}(\bm p_n) \approx \frac{\partial a_{m, k}}{\partial \tau_m}\nabla_{\bm p}\tau_m \nonumber\\
    &= -j\frac{2\pi}{\lambda_{k}} a_{m, k}(\bm p_n)\left( \frac{\bm p_n - \bm p_{{\rm Tx},m}}{\|\bm p_n - \bm p_{{\rm Tx},m} \|_2} + \frac{\bm p_n - \bm p_{{\rm Rx},m}}{\|\bm p_n - \bm p_{{\rm Rx},m} \|_2}\right).
\end{align}}\noindent
In 2D scenario, define the gradient reference channel gain matrices $\mathbf A_x = {\partial \mathbf A}/{\partial x} \in \mathbb{C}^{MK \times N}$ and $\mathbf A_y = {\partial \mathbf A}/{\partial y} \in \mathbb{C}^{MK \times N}$,
where 
{\small\begin{align}
    \frac{\partial \bm a_{m, k}}{\partial x}
    &=
    - j \frac{2\pi}{\lambda_{k}} \bm a_{m, k}(\bm p)\frac{x - x_{{\rm Tx},m}}{\|\bm p - \bm p_{{\rm Tx},m} \|_2} + \frac{x - x_{{\rm Rx},m}}{\|\bm p - \bm p_{{\rm Rx},m} \|_2}, \label{eq:ax}
    \\ 
    \frac{\partial \bm a_{m, k}}{\partial y}
    &=
    - j \frac{2\pi}{\lambda_{k}} \bm a_{m, k}(\bm p)\frac{y - y_{{\rm Tx},m}}{\|\bm p - \bm p_{{\rm Tx},m} \|_2} + \frac{y - y_{{\rm Rx},m}}{\|\bm p - \bm p_{{\rm Rx},m} \|_2}. \label{eq:ay}
\end{align}}\noindent
Therefore, the NLOS part of \eqref{eq:md} is approximated as
{\small \begin{align}
    \bm y^{\rm NLOS}  &= \mathbf \Phi\mathbf A(\bm p + \Delta \bm p)\bm x + \bm n \label{eq:ynlos1}\\
    &\approx \mathbf \Phi\big(\mathbf A(\bm p)\bm x + \nabla_{\bm p} \mathbf A(\bm p) (\Delta \bm p \circ \bm x)\big) + \bm n \label{eq:ynlos2}\\
    &= \mathbf \Phi\big(\mathbf A \bm x + \mathbf A_x \bm s_x + \mathbf A_y \bm s_y\big)  + \bm n \label{eq:ynlos3}\\
    &= \tilde {\mathbf A}\tilde {\bm x} + \bm n,\label{eq:ynlos}
\end{align}}\noindent
where $\tilde {\mathbf A} = \mathbf \Phi([\mathbf A, \mathbf A_x, \mathbf A_y]) \in \mathbb{C}^{MK \times 3N}$, $\tilde {\bm x}^{\mathsf T} = [\bm x^{\mathsf T}, \bm s^{\mathsf T}_x, \bm s^{\mathsf T}_y] \in \mathbb{R}^{3N}$ and $\circ$ represents Hadamard product.\footnote{The measurements in $\tilde{\mathbf A}$ are correlated, but the proposed method does not strictly depend on the irrelevance assumptions of $\tilde{\mathbf A}$.} Specifically, in \eqref{eq:ynlos2}, the first-order Taylor approximation as in \eqref{eq:Taylor} is adopted. In \eqref{eq:ynlos3}, we defined auxiliary variables $\bm s_x = \Delta \bm x \circ \bm x$ and $\bm s_y = \Delta \bm y \circ \bm x$ that incorporate the 2D off-grid offset. 

The resulting signal model reveals a strong coupling between the unknown environment parameters and the synchronization errors. Specifically, the measurement matrix depends on the PO matrix $\mathbf \Phi$, while accurate synchronization in turn relies on the reconstructed environment. This mutual dependency leads to a highly nonconvex and bilinear estimation problem for the scattering vector and the POs, which cannot be efficiently solved by conventional approaches \cite{tongx2}.

\begin{figure}[t]
    \centering
    \includegraphics[width=0.99\linewidth]{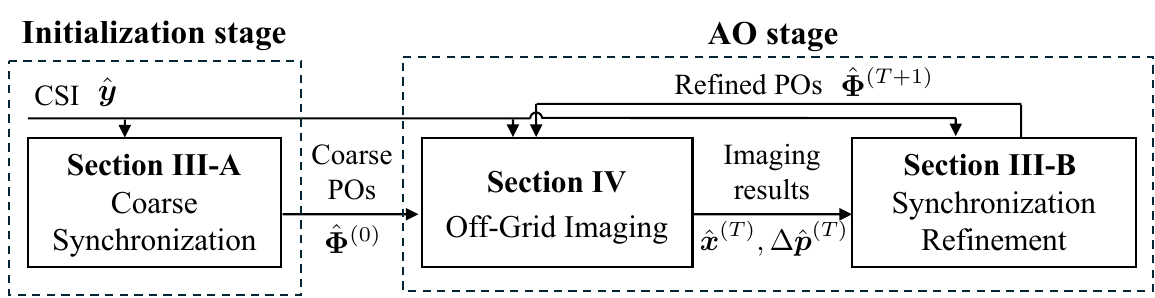}
    \caption{A sketch of the proposed joint OTA phase synchronization and off-Grid imaging method which iterates between imaging and synchronization during the AO stage. }
    \label{fig:sketch}
\end{figure}

To tackle this challenge, as shown in Fig.~\ref{fig:sketch}, we adopt a two-stage joint OTA phase synchronization and off-grid imaging method that combines coarse synchronization with iterative refinement. First, an OTA coarse synchronization is performed to provide a stable initialization for subsequent imaging algorithms, as detailed in Section~\ref{sec:coarsesyn}. 
Based on this initialization, we propose a joint AO framework. Let $T$ denote the iteration index. The algorithm alternates between the following two stages: \textit{(i) Off-Grid Imaging}: given the current PO matrix $\mathbf{\Phi}^{(T)}$, we estimate the environment sparse vector $\hat{\bm{x}}^{(T)}$ and off-grid offsets $\Delta \hat{\bm{p}}^{(T)}$ (Section~\ref{sec:imaging}); \textit{(ii) Synchronization Refinement}: using the estimated environment parameters from imaging stage as a reference, we update the PO matrix to $\mathbf{\Phi}^{(T+1)}$ (Section~\ref{sec:syn}). This process repeats until convergence. In the following sections, we describe the $T$-th AO iteration in detail.

\section{OTA Synchronization Algorithm}\label{sec:syn}
In this section, we propose a synchronization procedure consisting of a standard coarse step based on LOS components and an iterative fine synchronization stage, which forms the core contribution. The latter is jointly performed with the OG-AMP imaging algorithm in Section~\ref{sec:imaging} under an AO framework to achieve accurate imaging and phase synchronization.

\subsection{Coarse Synchronization}\label{sec:coarsesyn}
Before AO, to break the severe phase ambiguity inherent in distributed MIMO systems, this stage leverages the LOS path to initialize the synchronization parameters. 
The received channel frequency response (CFR) $\bm{y}_{m}$ for each Tx-Rx link is first transformed into the time domain via an inverse fast Fourier transform (IFFT), yielding a discrete-time channel impulse response. The corresponding power delay profile (PDP) is obtained as the squared magnitude of this time-domain response, i.e., $|{\rm IFFT}(\bm{y}_{m})|^{2}$, which characterizes the signal power distribution over propagation delays.
In this section, $\bm u$ is ignored for clarity. For links with a LOS component, we exploit the physical property that the LOS path always arrives earlier than NLOS reflections. Based on this observation, a time-domain gating operator $\mathcal{T}_{m}\{\cdot\}$ is applied to isolate the earliest significant delay tap corresponding to the LOS path. The extracted LOS component is then transformed back to the frequency domain as
\begin{equation} \small
    \hat{\bm h}_{m}^{\rm LOS} = {\rm FFT} \Big( \mathcal{T}_{m} \Big\{ {\rm IFFT}({\bm y}_{m}) \Big\} \Big).
\end{equation}
Using the estimated $\hat{h}_{m,k}^{\rm LOS}$ and the known ideal $h_{m,k}^{\rm LOS}$ in \eqref{eq:hLOS}, the coarse PO is calculated as
{\small \begin{align}
    \hat{\Phi}_{m,k} &= \angle \Big( \hat{h}_{m,k}^{\rm LOS} \cdot (h_{m,k}^{\rm LOS})^* \Big)\label{eq:phihat0}\\
    &\approx 2\pi (f_c + k \Delta f)\,\hat{\delta}_{t,m} + \phi_{{\rm LO},m},\label{eq:phihat1}
\end{align} }\noindent
In \eqref{eq:phihat0}, after identifying the dominant peak of the PDP, for example, via cross-correlation with the known pulse-shaping filter, we extract the initial phase $\hat{\Phi}_{m}$ directly from the corresponding complex observation rather than deriving it from the estimated TOA. The unknown residual phase error $\Delta \Phi_{m,k}$ caused by sampling error after this stage is expressed as the difference between the true and estimated phases,
{\small \begin{align}
\Delta \Phi_{m,k}&= \Phi_{m,k} - \hat{\Phi}_{m,k} \nonumber \\
&\approx 2\pi (f_c + k\Delta f)(\delta_{t,m}- \hat{\delta}_{t,m})
 - 2\pi f_c (\delta_{t,m}-\hat{\delta}_{t,m}) \nonumber \\
&= 2\pi k\Delta f (\delta_{t,m}-\hat{\delta}_{t,m}),
\label{eq:residual_phase_model}
\end{align}}\noindent
where $\hat{\delta}_{t,m}$ is the time delay corresponding to the estimated LOS path. This approach is inherently robust as it directly captures the large phase rotation $2\pi f_c \delta_{t,m}$ caused by the carrier frequency $f_c$ and the hardware-induced PO $\phi_{{\rm LO},m}$. After LOS-based coarse synchronization, residual POs may persist due to limited delay resolution caused by sampling errors and noise in PDP estimation.
The accuracy of $\hat{\Phi}_{m,k}$ is mainly constrained by two factors: (i) bandwidth-limited delay resolution, which may cause bias when NLOS paths are not resolvable from the LOS component, and (ii) SNR, which fundamentally limits the accuracy of both the PDP peak location and its complex amplitude. While zero-padding can reduce discretization effects, it cannot overcome these intrinsic limitations.
Nevertheless, since the LOS path typically has significantly higher power than NLOS components, the resulting residual phase error is expected to be moderate and provides a sufficiently coherent initialization for subsequent NLOS-aided imaging and fine synchronization. The estimation results in this section provide $\hat{\bf \Phi}^{(0)} = {\rm diag}([e^{-j\hat\Phi_{1,1}}, \dots, e^{-j\hat\Phi_{M,K}}]) \in \mathbb{C}^{MK \times MK}$ as a stable initialization for subsequent AO.

For links with a resolvable LOS path, i.e., $u_m=1$, a coarse estimate of the link-level phase $\Phi_{m,k}$ can be directly obtained from the LOS component. In contrast, for links with $u_m=0$, no direct phase observation is available at this stage. The available coarse phase observations can be written as
\begin{equation}\small
{\rm diag}(\tilde{\bm u}) \boldsymbol{\varphi}_k = {\rm diag}(\tilde{\bm u}) \mathbf G \boldsymbol{\phi}_k.
\end{equation}
Although links with $u_m=0$ do not provide direct phase measurements, their phases are indirectly constrained through shared Tx/Rx nodes. Specifically, once $\boldsymbol{\phi}_k$ is estimated from the available LOS links, the phases of all links can be obtained as $\hat \Phi_{m,k}=\mathbf g_m^{\mathsf T}\boldsymbol{\phi}_k$, the accuracy of which depends on the structure of $\mathbf G$.

\subsection{Synchronization Refinement}
After the LOS-aided coarse synchronization in Section~\ref{sec:coarsesyn} and the imaging stage in Section~\ref{sec:imaging}, a coarse estimate of sparse scattering coefficients $\hat{\bm{x}}^{(T)}$ and off-grid offsets $\Delta \hat{\bm{p}}_n^{(T)}$ obtained from estimates of $\bm s_x$ and $\bm s_y$ are available.
In this subsection, we exploit the reconstructed environment as a structural reference to refine the residual POs across distributed links. Different from conventional synchronization approaches, the proposed method performs OTA phase synchronization by leveraging the NLOS multipath components to refine the synchronization parameters for the $(T+1)$-th AO iteration.
We omit the superscript $T$ for variables within this section unless necessary.

The refined synchronization aims to align the observed CSI with the reconstructed NLOS response by compensating the residual POs.
As shown in Fig.~\ref{fig:sketch}, refined position $\hat{\bm p} = \bm p + \Delta \hat {\bm p}$ is calculated by adding the estimated off-grid offsets to the grid center.
We formulate the refined synchronization problem as maximum likelihood (ML) estimation
\begin{equation}\small
\hat{\bf \Phi} = \arg \min_{\bf \Phi} \|{\bf \Phi}^{-1}{\hat {\bm y}^{\rm NLOS}} - \tilde{\mathbf A}(\hat{\bm p}) \hat{\bm x} \|_2^2.\label{eq:phihat}
\end{equation}
Since $\bf \Phi$ is a diagonal matrix, optimization problem \eqref{eq:phihat} can be decomposed into the following ML analytical solution for the $m$-th link and $k$-th subcarrier, 
\begin{equation}\small
\hat{\Phi}_{m,k} = \angle\big(\hat y^{\rm NLOS}_{m,k} (h^{\rm NLOS}_{m,k})^* \big),
\label{eq:phihatmnf}
\end{equation}
where $h^{\rm NLOS}_{m,k} = \sum_{n=1}^N a_{m,k}(\hat{\bm p}_n)\hat x_n$.
According to the node-based phase model in \eqref{eq:association_matrix}, given the ML estimate $\hat{\bm \Phi}_{k}$ obtained from \eqref{eq:phihat}, the node-level phase vector is refined via least squares as
\begin{equation}\small
\hat{\bm \phi}_{k} = \big( \mathbf G^{\mathsf T} \mathbf G \big)^{-1}
\mathbf G^{\mathsf T}\hat{\bm \varphi}_{k},
\label{eq:node_ls}
\end{equation}
where $\hat{\boldsymbol \varphi}_{k} = [\hat\Phi_{1,k}, \ldots, \hat\Phi_{M,k}] \in \mathbb{C}^{M}$. The graph-consistent link phase is then reconstructed as
\begin{equation}\small
\tilde{\bm \varphi}_{k} = \mathbf G \hat{\bm \phi}_{k},
\end{equation}
which enforces node-level phase consistency while suppressing unstructured link-level noise.

Based on estimated POs, we update the matrix $\mathbf \Phi$ in \eqref{eq:md} for the $(T+1)$-th AO iteration,
\begin{equation}\small
\mathbf \Phi^{(T+1)} = \mathbf \Phi^{(T)} \cdot k_{\phi} \tilde{\bf\Phi},
\end{equation}
where $k_{\phi} \in (0,1)$ is the damping factor. Matrix $\tilde{\mathbf \Phi} = {\rm diag}([e^{-j\tilde{\bm \varphi}_{1}}, \dots, e^{-j\tilde{\bm \varphi}_{K}}]) \in \mathbb{C}^{MK \times MK}$. This updated matrix $\mathbf \Phi^{(T+1)}$ is then fed back into the OG-AMP algorithm in Section \ref{sec:imaging} to refine the sparse recovery, forming a closed-loop AO.

\section{OG-AMP Algorithm for Imaging}\label{sec:imaging}
In the $T$-th AO iteration, the PO matrix $\hat{\mathbf{\Phi}}^{(T)}$ is treated as known constant, we omit the superscript $T$ for variables within this section unless necessary. We employ the OG-AMP algorithm to solve for $\bm{x}$ and off-grid offsets given the observation $\bm{y}$ and fixed PO $\hat{\mathbf{\Phi}}$.
After applying the current synchronization, the ideal LOS component is removed from the estimated CSI. Based on \eqref{eq:ynlos}, the NLOS observation can be written as
{\small\begin{align}
\hat{\bm y}^{\text{NLOS}} &= \hat{\bf \Phi}^{-1}{\bm y} - \hat{\bm h}^{\rm LOS}\circ {\tilde{\bm u}} \nonumber \\
& = \Delta{\bf \Phi}\tilde{\bf A}\tilde{\bm x} + (\Delta{\bf \Phi} - {\bf I})\hat{\bm h}^{\rm LOS}\circ {\tilde{\bm u}} + \tilde{\bm n} \nonumber \\
& \approx \Delta{\bf \Phi}\tilde{\bf A}\tilde{\bm x} + \tilde{\bm n},\label{eq:yhat}
\end{align}}\noindent
where $\Delta{\bf \Phi} = {\rm diag}([e^{-j\Delta\Phi_{1,1}}, \dots, e^{-j\Delta\Phi_{M,K}}]) \in \mathbb{C}^{MK \times MK}$ denotes the unknown residual synchronization error matrix defined in \eqref{eq:residual_phase_model}. 
Vector $\tilde{\bm n}$ represents Gaussian white noise with variance $\sigma^2_{\rm n}$.
The term $(\Delta{\bf \Phi} - {\bf I}){\bm h^{\rm LOS}}$ represents the residual LOS leakage caused by imperfect synchronization. Since this term enters the model additively and is typically small after coarse synchronization, its impact on subsequent imaging performance can be absorbed into the noise.

In this section, we propose an OG-AMP algorithm for the off-grid model in Section~\ref{sec:OGC}, while existing works may alleviate modeling mismatch through pixel modeling, they fundamentally rely on on-grid models and do not explicitly resolve the off-grid nature of physical scatterers \cite{tongxicc, tongxtwc}. Furthermore, the proposed OG-AMP algorithm improves upon this by absorbing the residual phase synchronization error $\Delta\mathbf{\Phi}$ into the estimated complex amplitude and position offset. Even without explicit correction of these residual errors, the sparse structure of the path and the relative positions $\hat{\bm s}_x, \hat{\bm s}_y$ can still be approximately recovered. Therefore, OG-AMP exhibits robustness to PO and can assist in synchronization.

\subsection{Problem Formulation}
\begin{figure}[t]
    \centering
    \includegraphics[width=0.7\linewidth]{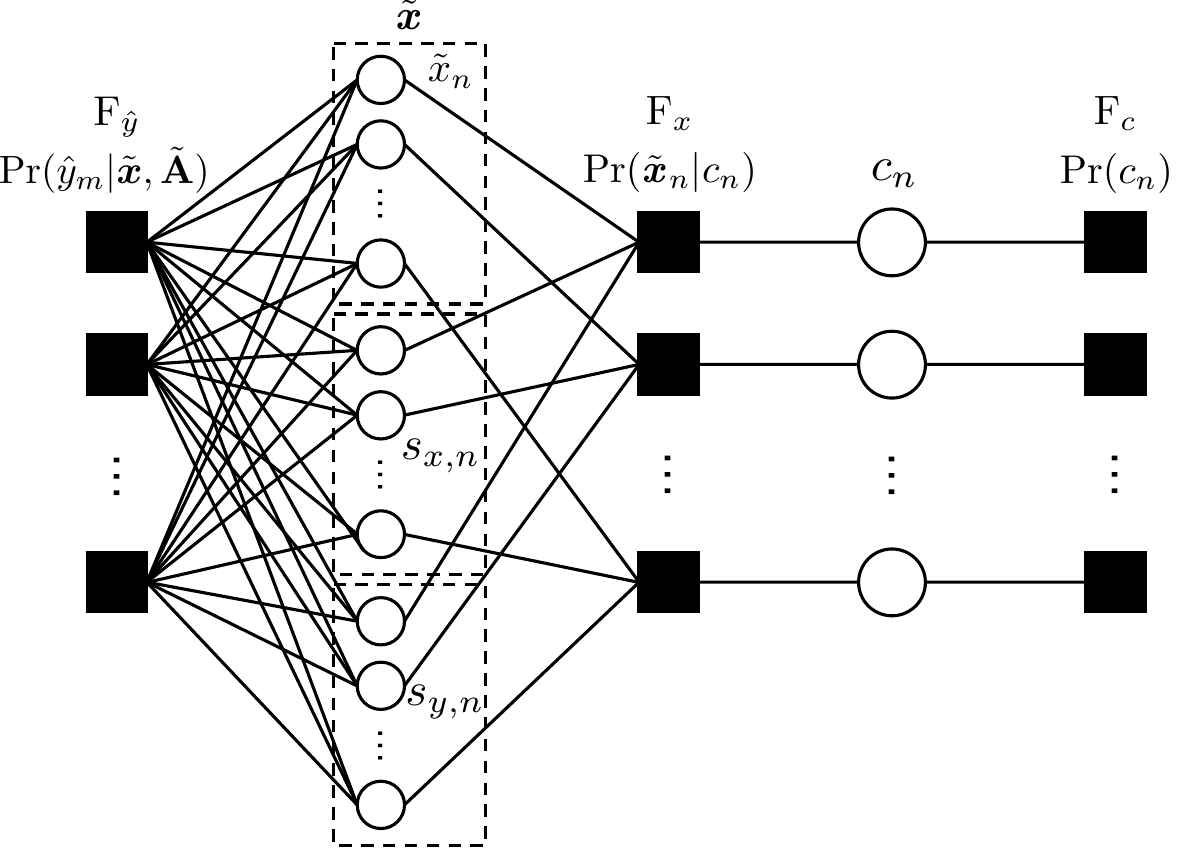}
    \caption{The factor graph of the OG-AMP algorithm, where squares represent FNs and circles represent VNs.}
    \label{fig:fg}
\end{figure}

We consider \eqref{eq:yhat} by ignoring the unknown residual PO estimation error $\Delta{\bf \Phi}$, the maximum\textit{ a posteriori} probability estimate of $\tilde {\bm x}$ is expressed as
\begin{equation}\small
    \hat {\bm x} = \arg \max_{\tilde{\bm x}} \Pr (\tilde{\bm x}|\hat{\bm y}, \tilde {\mathbf A}).
\end{equation}
Note that for clarity, the NLOS superscript of $\hat{\bm y}^{\rm NLOS}$ in Section~\ref{sec:imaging} is omitted. As shown in Fig.~\ref{fig:fg}, we propose a factor graph based on the decomposition result of the joint posterior probability,
\begin{equation}\label{eq:fg}\small
    \Pr (\tilde{\bm x}, \bm c|\hat{\bm y}, \tilde {\mathbf A}) \propto \Pr(\hat{\bm y}|\tilde{\bm x}, \tilde {\mathbf A})\Pr(\tilde{\bm x}|\bm c)\Pr(\bm c),
\end{equation}
where $\bm c \in \mathbb{B}^N$ is an activation index indicating that non-zero elements in $\bm x$, $\bm s_x$, and $\bm s_y$ have the same index. The goal is to compute the posterior probability of $c_n$ and the posterior means of $x_n$, $s_{x,n}$, and $s_{y,n}$.

We use a factor graph \cite{Kschischang} to represent the posterior probability in \eqref{eq:fg}. In Fig.~\ref{fig:fg}, circles represent variable nodes (VNs) $\tilde{\bm x}$ and $\bm c$, and squares represent factor nodes (FNs) $\Pr(\hat y_m|\tilde{\bm x}, \tilde {\mathbf A})$, $\Pr(\tilde{x}|c_n)$, and $\Pr(c_n)$, whose names are abbreviated as ${\rm F}_{\hat y}$, ${\rm F}_x$, and ${\rm F}_c$, respectively. The lines connecting the nodes indicate that these variables are included in the functions of the FNs. The specific functions of FNs are defined as follows:
\begin{itemize}
    \item FN ${\rm F}_c$: Define the sparse prior distribution of scatterers in the environment as
        % \begin{equation}
            $\Pr(c_n) = \eta^{c_n} (1-\eta)^{1-c_n},$
        % \end{equation}
        where $\eta$ is the sparsity, which needs to be roughly set in advance.

    \item FN ${\rm F}_x$: Define the distributions of variables ${x}_{n}$, $s_{x,n}$, and $x_{y,n}$ as
        \begin{equation}\small
        \Pr({x}_n|c_n) = \left\{
        \begin{aligned}
            &\mathcal{CN}(\mu_{x,n}, \sigma^2_{x,n}), \quad c_n = 1, \\
            &\delta({x}_n), \quad c_n = 0,
        \end{aligned}
        \right.
        \end{equation}
        \begin{equation}\small
            \Pr(s_{x,n}|c_n) = \left\{
            \begin{aligned}
                &\mathcal{CN}(\mu_{s_x,n}, \sigma^2_{s_x,n}), \quad c_n = 1, \\
                &\delta(s_x), \quad c_n = 0,
            \end{aligned}
            \right.
        \end{equation}
        \begin{equation}\small
            \Pr(s_{y,n}|c_n) = \left\{
            \begin{aligned}
                &\mathcal{CN}(\mu_{s_y,n}, \sigma^2_{s_y,n}), \quad c_n = 1, \\
                &\delta(s_y), \quad c_n = 0,
            \end{aligned}
            \right.
        \end{equation}
        where the variable $c_n$ controls whether variables ${x}_{n}$, $s_{x,n}$, and $x_{y,n}$ are activated together, with means $\mu_{x,n}$, $\mu_{s_x,n}$, and $\mu_{s_y,n}$, and variances $\sigma^2_{x,n}$, $\sigma^2_{s_x,n}$, and $\sigma^2_{s_y,n}$, respectively. Note that in the algorithm, we treat ${x}_{n}$ as a complex value, which absorbs the common offset caused by synchronization errors, thus enhancing the robustness of the algorithm.

        \item FN ${\rm F}_{\hat y}$: $\hat{\bm y}$ can be considered as an observation of $\tilde{\mathbf A}\tilde {\bm x}$ under Gaussian noise,
        % \begin{equation}
            $\Pr(\hat{\bm y}|\tilde{\bm x}, \tilde {\mathbf A}) = \mathcal{CN}(\tilde {\mathbf A}\tilde {\bm x}, \sigma^2_{\rm n}\mathbf{I}).$
        % \end{equation}
\end{itemize}

\subsection{Approximate Message Passing}\label{sec:AMP}
In this section, according to the factor graph, the messages passed iteratively between the VNs and FNs can be derived following the theory of sum-product algorithm \cite{Kschischang}. 
We follows a fundamental assumption of approximate message passing (AMP) theory \cite{Rangan}, we drop some high-order infinitesimal terms. Based on the central limit theorem, when $N \rightarrow \infty$, $MK/N$ remains constant, the messages are approximated as Gaussian random variables, thereby calculating the approximate marginal posterior distribution of the unknown variables. Note that all $x_n$, $s_{x,n}$, and $s_{y,n}$ below are equivalent and their derivations are similar.

\subsubsection{From FN ${\rm F}_{\hat y}$ to VN $\tilde{\bm x}$}
As derived in Appendix~\ref{ap:A}, the mean and variance of the message $\mu_{{\rm F}_{\hat y,m}\rightarrow x_{n}}^{(t)}(x_n)$ in $t$-th iteration is updated as
\begin{equation}\small
    \hat x_{{m} \rightarrow x_n}^{(t)} = \frac{\hat y_m - Z_{{m} \rightarrow x_n}^{(t)}}{a_{m,n}},
    \label{eq:xmtxn}
\end{equation}
\begin{equation}\small
    v_{{m} \rightarrow x_n}^{(t)} = \frac{\sigma_{\rm n}^2 + V_{m \rightarrow x_n}^{(t)}}{|a_{m,n}|^2}.
    \label{eq:vmtxn}
\end{equation}
The derivations of ${s}_{x,n}$ and ${s}_{y,n}$ are similar and will not be repeated here. 

\subsubsection{From VN $\tilde{\bm x}$ to FN ${\rm F}_x$}
As derived in Appendix~\ref{ap:B}, the mean and variance of the message $\mu_{{x_n} \rightarrow {\rm F}_{x,n}}^{(t)}(x_n)$ in $t$-th iteration is updated as
{\small \begin{align}
\hat{x}_{n \to {\rm F}_x}^{(t)} &= v_{n \to {\rm F}_x}^{(t)} \Big( \sum_{m=1}^{M} \frac{\hat{x}_{m\rightarrow x_{n}}^{(t)}}{v_{m\rightarrow x_{n}}^{(t)}} \Big),
\label{eq:xntfx}
\end{align}}\noindent
{\small \begin{align}
v_{n \to {\rm F}_x}^{(t)} &= \Big( \sum_{m=1}^{M} \frac{1}{v_{m\rightarrow x_{n}}^{(t)}} \Big)^{-1}.
\label{eq:vntfx}
\end{align}}\noindent
The derivations of ${s}_{x,n}$ and ${s}_{y,n}$ are similar and will not be repeated here. 
\subsubsection{From FN ${\rm F}_x$ to VN $\textbf{\textit{c}}$}
The message $\mu^{(t)}_{{\rm F}_{x,n} \rightarrow {c_n}}(c_n)$ in $t$-th iteration is expressed as 
\begin{equation}
    \mu^{(t)}_{{\rm F}_{x,n} \rightarrow {c_n}}(c_n) \propto Z_{x,n}^{(t)}(c_n)Z_{s_{x,n}}^{(t)}(c_n)Z_{s_{y,n}}^{(t)}(c_n),
    \label{eq:mfymtcn}
\end{equation}
where $Z_{x,n}^{(t)}(c_n) = \int \Pr(x_n|c_n) \mu_{x_{n}\rightarrow {\rm F}_{x,n}}^{(t)}(x_n) {\rm d}x_n $. The definitions of $Z_{s_{x,n}}^{(t)}(c_n)$ and $Z_{s_{y,n}}^{(t)}(c_n)$ are similar and will not be repeated here. As derived in Appendix~\ref{ap:C}, $Z_{x,n}^{(t)}(c_n)$ is calculated as
{\small\begin{align}
    Z_{x,n}^{(t)}(0) &= \mathcal{CN}(0 ; \hat{x}_{n\rightarrow {\rm F}_x}^{(t)}, v_{n\rightarrow {\rm F}_x}^{(t)}), \label{eq:zxnt0}\\
    Z_{x,n}^{(t)}(1) &= \mathcal{CN}(\mu_{x,n} ; \hat{x}_{n\rightarrow {\rm F}_x}^{(t)}, \sigma_{x,n}^2 + v_{n\rightarrow {\rm F}_x}^{(t)}).\label{eq:zxnt1}
\end{align}}\noindent
Multiplying the results of $Z_{x,n}^{(t)}(c_n)$, $Z_{s_{x,n}}^{(t)}(c_n)$ and $Z_{s_{y,n}}^{(t)}(c_n)$, we obtain the message passed to $c_n$.

\subsubsection{From FN ${\rm F}_c$ to VN $\textbf{\textit{c}}$}
This message is unrelated to iteration, $\mu_{{\rm F}_{c,n} \rightarrow c_n}(c_n) = \Pr(c_n)$.

\subsubsection{From VN $\textbf{\textit{c}}$ to FN ${\rm F}_x$}
This message is unrelated to iteration, $\mu_{c_n \rightarrow {\rm F}_{x,n}}(c_n) \propto \mu_{{\rm F}_{c,n} \rightarrow c_n}(c_n)$.

\subsubsection{From FN ${\rm F}_x$ to VN $\tilde{\bm x}$}
As derived in Appendix~\ref{ap:D}, the message $\mu_{{\rm F}_{x,n}\rightarrow x_{n}}^{(t+1)}(x_{n})$ 
% in $(t+1)$-th iteration 
is approximated to
\begin{equation}\small
    \mu_{{\rm F}_{x,n}\rightarrow x_{n}}^{(t+1)}(x_{n}) \propto k_{n,0}^{(t+1)}  \delta(x_n) + k_{n,1}^{(t+1)} \mathcal{CN}(x_{n};\mu_{x,n},\sigma_{x,n}^{2}),
    \label{eq:fxntxn1}
\end{equation}
where weights $k_{n,0}^{(t+1)}$ and $k_{n,1}^{(t+1)}$ are defined as
\begin{equation}\small
    k_{n,0}^{(t+1)} \propto (1-\eta) \cdot Z_{s_{x,n}}^{(t)}(0) \cdot Z_{s_{y,n}}^{(t)}(0),
    \label{eq:kn0}
\end{equation}
\begin{equation}\small
    k_{n,1}^{(t+1)} \propto \eta \cdot Z_{s_{x,n}}^{(t)}(1) \cdot Z_{s_{y,n}}^{(t)}(1).
    \label{eq:kn1}
\end{equation}
The messages $\mu_{{\rm F}_{x,n}\rightarrow s_{x,n}}^{(t+1)}$ and $\mu_{{\rm F}_{x,n}\rightarrow s_{y,n}}^{(t+1)}$ passed to $s_{x,n}$ and $s_{y,n}$ also have the exact same form, only the corresponding weights need to be adjusted accordingly.

\subsubsection{From VN $\tilde{\bm x}$ to FN ${\rm F}_{\hat y}$}
As derived in Appendix~\ref{ap:E}, the mean and variance in \eqref{eq:zmtxn} and \eqref{eq:vvmtxn} of the message $\mu_{x_n \rightarrow {\rm F}_{\hat y,m}}^{(t+1)}(x_n)$ are approximated to
{\small\begin{align}
    \hat{x}_{n \to {\rm F}_{\hat y,m}}^{(t+1)}= P_{n,1 \to m}^{(t+1)} \cdot \hat{x}_{{\rm prod}, m}^{(t+1)}, 
    \label{eq:xntfym}
\end{align}}\noindent
{\small \begin{align}
    v_{n \to {\rm F}_{\hat y,m}}^{(t+1)} &= P_{n,1 \to m}^{(t+1)} v_{{\rm prod}, m}^{(t+1)}  \nonumber \\ &+ \left( P_{n,1 \to m}^{(t+1)} - (P_{n,1 \to m}^{(t+1)})^2 \right) |\hat{x}_{{\rm prod}, m}^{(t+1)}|^2,
    \label{eq:vntfym}
\end{align}}\noindent
where the normalized weights $P_{n,1 \to m}^{(t+1)}$ other intermediate variables are defined in Appendix~\ref{ap:E}.
The derivations of ${s}_{x,n}$ and ${s}_{y,n}$ are similar and will not be repeated here. 

\subsection{Estimation and Iterative Execution}
\subsubsection{Estimation for Active Pixels}
In $t$-th iteration, for each $n$, calculate the posterior probability of $c_n$ as
{\small \begin{align}
    \Pr(c_n=1|\hat{\bm y},\tilde{\bf A}) &\propto \mu_{{\rm F}_{c,n} \to c_n}(1) \cdot \mu_{{\rm F}_{x,n} \to c_n}^{(t)}(1) \\
    &\propto \eta \cdot Z_{x,n}^{(t)}(1) Z_{s_{x,n}}^{(t)}(1) Z_{s_{y,n}}^{(t)}(1),
    \label{eq:pcn1}
\end{align}}\noindent
{\small \begin{align}
    \Pr(c_n=0|\hat{\bm y},\tilde{\bf A}) &\propto \mu_{{\rm F}_{c,n} \to c_n}(0) \cdot \mu_{{\rm F}_{x,n} \to c_n}^{(t)}(0) \\
    &\propto (1-\eta) \cdot Z_{x,n}^{(t)}(0) Z_{s_{x,n}}^{(t)}(0) Z_{s_{y,n}}^{(t)}(0).
    \label{eq:pcn0}
\end{align}}\noindent
If $\Pr(c_n=1|\hat{\bm y},\tilde{\bf A}) / \big(\Pr(c_n=1|\hat{\bm y},\tilde{\bf A})+\Pr(c_n=0|\hat{\bm y},\tilde{\bf A})\big)  > \eta_c $, set $\hat{c}_n^{(t)} = 1$. Otherwise, set $\hat{c}_n^{(t)} = 0$. Generally, we set $\eta_c=0.5$. The set of active pixels in $t$-th iteration is defined as $\mathcal{A}^{(t)} = \{n | \hat{c}_n^{(t)} = 1\}$.

In this step, a key aspect of the proposed algorithm is that matrices $\mathbf A$, $\mathbf A_x$ and $\mathbf A_y$ are not static during iteration. They are recalculated in each iteration $t$ based on the estimated offset position. At the start of the $t$-th iteration, we use the offsets $\hat{s}_{x,n}^{(t-1)}$ and $\hat{s}_{y,n}^{(t-1)}$ estimated in the previous $(t-1)$-th iteration to calculate the gradient matrices $\mathbf A_x^{(t)}$ and $\mathbf A_y^{(t)}$, which effectively reduces the error of the first-order Taylor approximation in \eqref{eq:Taylor}.

\subsubsection{Estimation for Scattering Coefficient and Off-Grid Pixel offsets}
For inactive pixels $n \notin \mathcal{A}$: $\hat{x}_n^{(t)} = 0$, $\hat{s}_{x,n}^{(t)} = 0$, $\hat{s}_{y,n}^{(t)} = 0$. For active pixels $n \in \mathcal{A}^{(t)}$, 
% \begin{align}
    $\Pr(x_n|{\hat{\bm y}}, \tilde{\bf A}) \propto \mu_{{\rm F}_{x,n}\rightarrow x_{n}}(x_n) \cdot \prod_{m=1}^{M} \mu_{{\rm F}_{\hat y,m} \to x_n}^{(t)}(x_n) \propto \mathcal{CN}(x_{n};\mu_{x,n},\sigma_{x,n}^{2})\mathcal{CN}(x_{n};\hat{x}_{n\rightarrow {\rm F}_{x}}^{(t)},v_{n\rightarrow {\rm F}_{x}}^{(t)}).$
% \end{align}
The estimated mean $\hat{x}_n^{(t)}$ and variance $v_n^{(t)}$ are
\begin{equation}\small
    \hat{x}_n^{(t)} = v_n^{(t)} \left( \frac{\mu_{x,n}}{\sigma_{x,n}^2} + \frac{\hat{x}_{n\rightarrow {\rm F}_{x}}^{(t)}}{v_{n\rightarrow {\rm F}_{x}}^{(t)}} \right),
    \label{eq:xn}
\end{equation}
\begin{equation}\small
    v_n^{(t)} = \left( \frac{1}{\sigma_{x,n}^2} + \frac{1}{v_{n\rightarrow {\rm F}_{x}}^{(t)}} \right)^{-1}.
    \label{eq:vn}
\end{equation}
The derivations of $\hat{s}_{x,n}^{(t)}$ and $\hat{s}_{y,n}^{(t)}$ are similar and will not be repeated here. 

\begin{breakablealgorithm}
  \caption{\vspace{-0.0cm}The proposed OG-AMP Algorithm}
  \label{alg:OG-AMP}
  \begin{algorithmic}[1]
  \REQUIRE
  The frequency $f_{k}$. The positions $\bm p_{\rm Tx}$, $\bm p_{\rm Rx}$ and $\bm p$. NLOS channel estimation results $\hat{\bm y}$. Prior probability parameters $\mu_{x,n}$, $\mu_{s_x,n}$, $\mu_{s_y,n}$, $\sigma^2_{x,n}$, $\sigma^2_{s_x,n}$, and $\sigma^2_{s_y,n}$.
  \STATE
  \textbf{Initialization}: Set $\hat{x}_{n\rightarrow {\rm F}_{\hat y,m}}^{(0)} = \mu_{x,n}$, $v_{n\rightarrow {\rm F}_{\hat y,m}}^{(0)} = \sigma_{x,n}^2$. Set $\hat{s}_{x, n\rightarrow {\rm F}_{\hat y,m}}^{(0)}$, $v_{s_x, n\rightarrow {\rm F}_{\hat y,m}}^{(0)}$, $\hat{s}_{y, n\rightarrow {\rm F}_{\hat y,m}}^{(0)}$, $v_{s_y, n\rightarrow {\rm F}_{\hat y,m}}^{(0)}$ similarly. Initialize $\tilde{\mathbf{A}}^{(0)}$ according to \eqref{eq:fr}, \eqref{eq:ax} and \eqref{eq:ay}.

  \WHILE {$t<t_{\rm max}$ and $\|\hat{\bm x}^{(t)} - \hat{\bm x}^{(t-1)}\| / M > \varepsilon$, where $\varepsilon$ is a given error tolerance value}
  \STATE
    \textit{Update Messages from FN ${\rm F}_{\hat y} \to$ VN $\tilde{\bm x}$}:
    \begin{itemize}[leftmargin=*, topsep=0pt, itemsep=0pt, parsep=0pt, partopsep=0pt]
        \item Calculate $Z_{m\rightarrow x_{n}}^{(t)}$ and $V_{m\rightarrow x_{n}}^{(t)}$ according to \eqref{eq:zmtxn} and \eqref{eq:vvmtxn}. Repeat for $Z_{m\rightarrow s_{x,n}}^{(t)}$, $V_{m\rightarrow s_{x,n}}^{(t)}$, $Z_{m\rightarrow s_{y,n}}^{(t)}$, $V_{m\rightarrow s_{y,n}}^{(t)}$.

        \item Calculate $\hat x_{{m} \rightarrow x_n}^{(t)}$ and $v_{{m} \rightarrow x_n}^{(t)}$ according to \eqref{eq:xmtxn} and \eqref{eq:vmtxn}. Repeat for $\hat{s}_{x, m\rightarrow x_{n}}^{(t)}$, $v_{s_x, m\rightarrow x_{n}}^{(t)}$, $\hat{s}_{y, m\rightarrow x_{n}}^{(t)}$, $v_{s_y, m\rightarrow x_{n}}^{(t)}$.
    \end{itemize}
  \STATE
    \textit{Update Messages From VN $\tilde{\bm x} \to$ FN ${\rm F}_x$}:
    \begin{itemize}[leftmargin=*, topsep=0pt, itemsep=0pt, parsep=0pt, partopsep=0pt]
        \item Calculate $\hat{x}_{n \to {\rm F}_x}^{(t)}$ and $\hat{v}_{n \to {\rm F}_x}^{(t)}$ according to \eqref{eq:xntfx} and \eqref{eq:vntfx}. 

        \item Repeat for $\hat{s}_{x,n\rightarrow {\rm F}_{x}}^{(t)}$, $ v_{s_x,n\rightarrow {\rm F}_{x}}^{(t)}$, $\hat{s}_{y,n\rightarrow {\rm F}_{x}}^{(t)}$, $v_{s_y,n\rightarrow {\rm F}_{x}}^{(t)}$.
    \end{itemize}
  \STATE
    \textit{Update Messages From FN ${\rm F}_x \to$ VN $\tilde{\bm x} \to$ FN ${\rm F}_{\hat y}$}:
    \begin{itemize}[leftmargin=*, topsep=0pt, itemsep=0pt, parsep=0pt, partopsep=0pt]
        \item Calculate terms $Z_{x,n}^{(t)}(0)$ and $Z_{x,n}^{(t)}(1)$ according to \eqref{eq:zxnt0} and \eqref{eq:zxnt1}. Repeat for $Z_{s_{x},n}^{(t)}(0)$, $Z_{s_{x},n}^{(t)}(1)$, $Z_{s_{y},n}^{(t)}(0)$, $Z_{s_{y},n}^{(t)}(1)$.
        \item Calculate parameters $\hat{x}_{{\rm cav}, m}^{(t)}$ and $v_{{\rm cav}, m}^{(t)}$ according to \eqref{eq:xcav} and \eqref{eq:vcav}. Repeat for $s_x$ and $s_y$.
        \item Calculate parameters $\hat{x}_{{\rm prod}, m}^{(t+1)}$ and $v_{{\rm prod}, m}^{(t+1)}$ according to \eqref{eq:xprod} and \eqref{eq:vprod}. Repeat for $s_x$ and $s_y$.
        \item Calculate weights $k_{n,0}^{(t+1)}$, $k_{n,1}^{(t+1)}$, $W_{n,0}^{(t+1)}$, $W_{n,1}^{(t+1)}$ and $P_{n,1 \to m}^{(t+1)}$ according to \eqref{eq:kn0}, \eqref{eq:kn1}, \eqref{eq:wn0}, \eqref{eq:wn1} and \eqref{eq:pn1}. Repeat for weights of $s_x, s_y$.
        \item Calculate $\hat{x}_{n\rightarrow {\rm F}_{\hat y,m}}^{(t+1)}$ and $v_{n\rightarrow {\rm F}_{\hat y,m}}^{(t+1)}$ according to \eqref{eq:xntfym} and \eqref{eq:vntfym}. Repeat for $\hat{s}_{x,n\rightarrow {\rm F}_{\hat y,m}}^{(t+1)}$, $v_{s_x,n\rightarrow {\rm F}_{\hat y,m}}^{(t+1)}$, $\hat{s}_{y,n\rightarrow {\rm F}_{\hat y,m}}^{(t+1)}$, $v_{s_y,n\rightarrow {\rm F}_{\hat y,m}}^{(t+1)}$.
    \end{itemize}
  \STATE
    \textit{Estimation and Gradients Update}:
    \begin{itemize}[leftmargin=*, topsep=0pt, itemsep=0pt, parsep=0pt, partopsep=0pt]
        \item Estimate $\mathcal{A}^{(t)}$ according to \eqref{eq:pcn1} and \eqref{eq:pcn0}.
        \item For $n \notin \mathcal{A}^{(t)}$: $\hat{x}_n^{(t)} = 0$, $\hat{s}_{x,n}^{(t)} = 0$, $\hat{s}_{y,n}^{(t)} = 0$. Keep the gradients unchanged: $\mathbf A^{(t+1)}(:, n) = \mathbf A^{(t)}(:, n)$, $\mathbf A_x^{(t+1)}(:, n) = \mathbf A_x^{(t)}(:, n)$, $\mathbf A_y^{(t+1)}(:, n) = \mathbf A_y^{(t)}(:, n)$.
        \item For $n \in \mathcal{A}(t)$: Estimate $\hat{x}_{n}^{\rm final}$ and $v_{n}^{\rm final}$ according to \eqref{eq:xn} and \eqref{eq:vn}. Repeat for $\hat{s}_{x,n}^{(t)}$ and $\hat{s}_{y,n}^{(t)}$. Calculate the shift $\Delta\hat{x}_{n}^{(t)} = \hat{s}_{x,n}^{(t)}/\hat{x}_{n}^{(t)}$ and repeat for $\Delta\hat{y}_{n}^{(t)}$.  
        Update $\mathbf A^{(t+1)}(:, n)$ and the gradients $\mathbf A_{x}^{(t+1)}(:, n)$ and $\mathbf A_{y}^{(t+1)}(:, n)$ based on the offset position according to \eqref{eq:ax} and \eqref{eq:ay}.
    \end{itemize}
  \STATE
  Set $t = t + 1$. In each iteration, execute steps~3-5. Every $t'$-th iteration, also execute step~6.
  \ENDWHILE
  \ENSURE
  Scattering coefficient $\hat{x}_n^{(t)}$ and offsets $\Delta\hat{x}_{n}^{(t)}$, $\Delta\hat{y}_{n}^{(t)}$.
  \end{algorithmic}
  \end{breakablealgorithm}

\subsubsection{Iterative Execution}
The message passing rules derived in Section~\ref{sec:AMP} define an iterative algorithm. We summarize the proposed OG-AMP algorithm in Algorithm~\ref{alg:OG-AMP}. 
In each iteration of the OG-AMP algorithm, the most computationally intensive operations are matrix-vector multiplications between the measurement matrix $\tilde{\mathbf A}$ and the message vectors. Specifically, the update of messages in step 3 and step 4 requires $\mathcal{O}(MN)$ operations. The calculation of posterior statistics and weights (step 5 and 6) involves element-wise operations with linear complexity $\mathcal{O}(N)$ or $\mathcal{O}(M)$. Consequently, the overall complexity of the imaging stage is $\mathcal{O}(MNK)$. Therefore, the total complexity scales linearly with the number of subcarriers and the pixel number, making the method computationally efficient for large-scale distributed MIMO systems compared to algorithms requiring matrix inversions, such as minimum mean squared error (MMSE) estimation, which scale as $\mathcal{O}((3N)^3)$.

\section{Theoretical Analysis}\label{sec:per}
In this section, we analyze the system performance under the ideal setting and the practical scenario with synchronization and off-grid errors.
\subsection{Imaging Performance for Perfect Synchronization}\label{sec:perA}
Consider the ideal scenario with perfect synchronization ($\mathbf \Phi=\mathbf I$) and no off-grid offsets $\Delta \bm p = \mathbf 0$, we will demonstrate the impact of basic parameter settings on system performance.
The imaging model reduces to $\bm y = \mathbf A\bm x + \bm n$, where $\mathbf A\in\mathbb{C}^{M K\times N}$ is the measurement matrix and $\bm x$ is an $s$-sparse vector. Although the measurement matrix $\mathbf A$ is structured and not i.i.d. Gaussian, it has been widely observed in ISAC, radar imaging, and compressed sensing literature that the columns of $\mathbf A$ behave approximately uncorrelated when the number of measurements $MK$ is large. Stable sparse recovery is guaranteed when the number of measurements satisfies the classical compressed sensing scaling
$M K \geq Cs\log(N/s)$ where $C$ is a constant which depends on the RIP level of $\mathbf A$, as derived in \cite{Candes,baraniuk2008simple,tongx2}.
Under this condition, the reconstruction error is expressed as
\begin{equation}\small
    {\rm E}\left[\|\hat{{\bm x}}-{\bm x}\|_2^2\right] \propto \sigma_{\rm n}^2\, \frac{\eta\log N}{MK}.
    \label{eq:mse_scaling_explicit}
\end{equation}
Eq.~\eqref{eq:mse_scaling_explicit} serves as an ideal benchmark for the subsequent analysis. It indicates that increasing the number of links $M$ or subcarriers $K$ improves reconstruction accuracy, while higher sparsity level and noise variance $\sigma_{\rm n}^2$ lead to increased estimation error.

\subsection{Synchronization Performance Bound for Perfect Imaging}\label{sec:perB}
In the proposed method, residual POs are modeled through a node-based phase estimation.
After LOS-aided coarse synchronization and imaging, the refined synchronization stage estimates the link-level phases and then projects them onto the node-consistent subspace.
In this subsection, we analyze the performance bound of estimating the node-level phase parameters $\bm \phi_{k}$ and discuss their impact on imaging performance.

\subsubsection{CRB for PO Estimation}
Assuming perfect imaging, the received CSI for link $m$ and subcarrier $k$ is
% \begin{equation}
$y_{m,k}
= e^{-j\phi_{m,k}} h_{m,k} + n_{m,k},$
% \end{equation}
where $h_{m,k}$ is the ideal channel coefficient and $n_{m,k} \sim \mathcal{CN}(0,\sigma_{\rm n}^2)$.
Using the sufficient statistic
% \begin{align}
$z_{m,k} = \angle\big( y_{m,k} (h_{m,k})^* \big) = \phi_{m,k} + w_{m,k},$
% \end{align}
the effective phase noise $w_{m,k}$ can be approximated at high SNR as
\begin{equation}\small
\mathrm{Var}(w_{m,k}) \approx \frac{\sigma_{\rm n}^2}{2|h_{m,k} |^2}.
\end{equation}
Stacking all $M$ links at subcarrier $k$, we obtain the linear Gaussian model
% \begin{equation}
$\bm z_{k}
= \mathbf G \bm \phi_{k} + \bm \eta_{k},$
% \end{equation}
where $\bm \eta_{k}$ has diagonal covariance matrix $\mathbf \Sigma_{k} = \mathrm{diag}( \sigma_{m,k}^2 )$ with $\sigma_{m,k}^2 = \mathrm{Var}(w_{m,k})$. The Fisher information matrix for $\bm \phi_{k}$ is therefore given by $
\mathbf J_{k}
= \mathbf G^{\mathsf T} \mathbf \Sigma_{k}^{-1} \mathbf G$, and the corresponding CRB is $\mathbf J_{k}^{-1}$.
To reveal the impact of NLOS propagation, note that
\begin{equation}\small
|h_{m,k}|^2
= \Big|\sum_{n \in \mathcal A}a_{m,k}(\bm p_n + \Delta {\bm p}_n) x_n\Big|^2.
\end{equation}
Therefore, the CRB is approximated as
\begin{equation}\small
\mathrm{CRB}(\bm \phi_k)
\propto
\frac{\sigma_{\rm n}^2}{|\mathcal A|}
\left(\mathbf G^{\mathsf T}\mathbf G\right)^{-1}.
\end{equation}
It can be concluded that the estimation accuracy of node-level phase parameters is jointly determined by the network topology encoded in $\mathbf G$, the effective signal strength of reconstructed NLOS components $|\mathcal A|$, and the presence of residual phase errors. 
A well-connected network topology improves the conditioning of $\mathbf G^{\mathsf T}\mathbf G$, thereby significantly reducing the CRB. Moreover, spatially diverse NLOS multipath components enhance the effective SNR of phase observations, which further improves synchronization accuracy. This analysis explains why exploiting NLOS multipath as a structural reference is essential for achieving reliable OTA synchronization in distributed MIMO ISAC systems.

\subsubsection{Impact on Imaging Performance}
After node-level phase refinement, the residual estimation error $\Delta \bm \phi_{k} = \bm \phi_{k} - \hat{\bm \phi}_{k}$ induces a structured residual phase error on each link,
% \begin{equation}
$\Delta \bm \varphi_{k} = \mathbf G \Delta \bm \phi_{k}.$
% \end{equation}
For small residual errors, the resulting multiplicative distortion can be linearized as
% \begin{equation}
$e^{-j\Delta \varphi_{m,k}} \approx 1 - j \Delta \varphi_{m,k},$
% \end{equation}
which converts the phase error into an additive perturbation on the measurement matrix.
The equivalent noise variance induced by node-level synchronization errors is expressed as
{\small \begin{align}
\sigma_{\rm syn}^2 & \approx  \mathrm{E}\left[ (\Delta \varphi_{m,k})^2 \right] |h_{m,k}|^2 \\
& = |h_{m,k}|^2
\left[ \mathbf G \mathbf{J}^{-1}_k \mathbf G^{\mathsf T} \right]_{m,m} \\
& \propto \frac{\sigma_{\rm n}^2}{|\mathcal A|}|h_{m,k}|^2
\left[\mathbf G(\mathbf G^{\mathsf T}\mathbf G)^{-1}\mathbf G^{\mathsf T}\right]_{m,m}.
\end{align}}\noindent
This expression reveals how node-level phase uncertainty propagates to link-level distortions and degrades the measurement SNR. As a result, synchronization accuracy,
network topology, and the number of available subcarriers jointly determine the imaging performance.

\subsection{Off-Grid Approximation Error}
In this subsection, we will analyze the impact of residuals on performance caused by the linear approximation of the off-grid offset. The second-order remainder in the first-order Taylor expansion \eqref{eq:Taylor} is given by 
\begin{equation}\small
\varepsilon_{m,k}^{\text{off}} = \frac{1}{2} \Delta\bm{p}_n^\mathsf{T} \, \nabla_{\bm{p}}^2 a_{m,k}(\bm{p}_n) \, \Delta\bm{p}_n,
\end{equation}
For propagation over frequency $f_{k}$, we obtain 
% \begin{equation}
$\nabla_{\bm p}^2 a_{m,k}\sim \Big(\frac{2\pi}{\lambda_k}\Big)^2 a_{m,k}.$
% \end{equation}
Substituting into the remainder, we give the off-grid error as
\begin{equation}\small
{\rm E}\big[|\varepsilon_{m,k}^{\rm off}|^2\big]
    \propto
    \Big(\frac{2\pi}{\lambda_k}\Big)^{4}{\rm E}\left[\|\Delta {\bm p}_n\|_2^{4}\right].
    \label{eq:off_scaling_basic}
\end{equation}
With total subcarriers $K$, the total off-grid distortion is expressed as
\begin{equation}\small
    \sigma_{\rm off}^2 \propto \frac{1}{K} \sum_{k=1}^{K} 
    \Big(\frac{2\pi}{\lambda_k}\Big)^{4} {\rm E}[\|\Delta {\bm p}_n\|^{4}_2].
\end{equation}
After polynomial expansion, the dominant term is governed by the carrier wavelength $\lambda_c$, and we approximate $\lambda_k^4 \approx \lambda_c^4$.
Assuming that the off-grid displacement $\Delta \bm p_n$ is uniformly
distributed within a square pixel of side length $d_{\mathrm{pixel}}$,
the resulting variance of the off-grid approximation error can be expressed as
\begin{equation}\small
    \sigma_{\mathrm{off}}^{2}
    \propto
    \frac{1}{K}
    \left(\frac{1}{\lambda_c}\right)^{4}
    d_{\mathrm{pixel}}^{4}.
\end{equation}
We conclude that the off-grid error increases with $\lambda^{-4}$ and the fourth power of the pixel size, while decreasing with the number of subcarriers. This implies a practical resolution constraint that the pixel size should remain in the order of a few wavelengths to avoid excessive phase offsets. However, our method also outperforms conventional methods \cite{tongx, tongx2} in this respect due to its robustness under phase error conditions.

\begin{figure*}[t]
  \centering
  \begin{minipage}[t]{0.22\textwidth}
      \centering
      \subfigure[On-grid imaging with GAMP, no PO. (${\rm CD} = 0.0375~\rm m$)]{
      \includegraphics[width=0.95\textwidth]{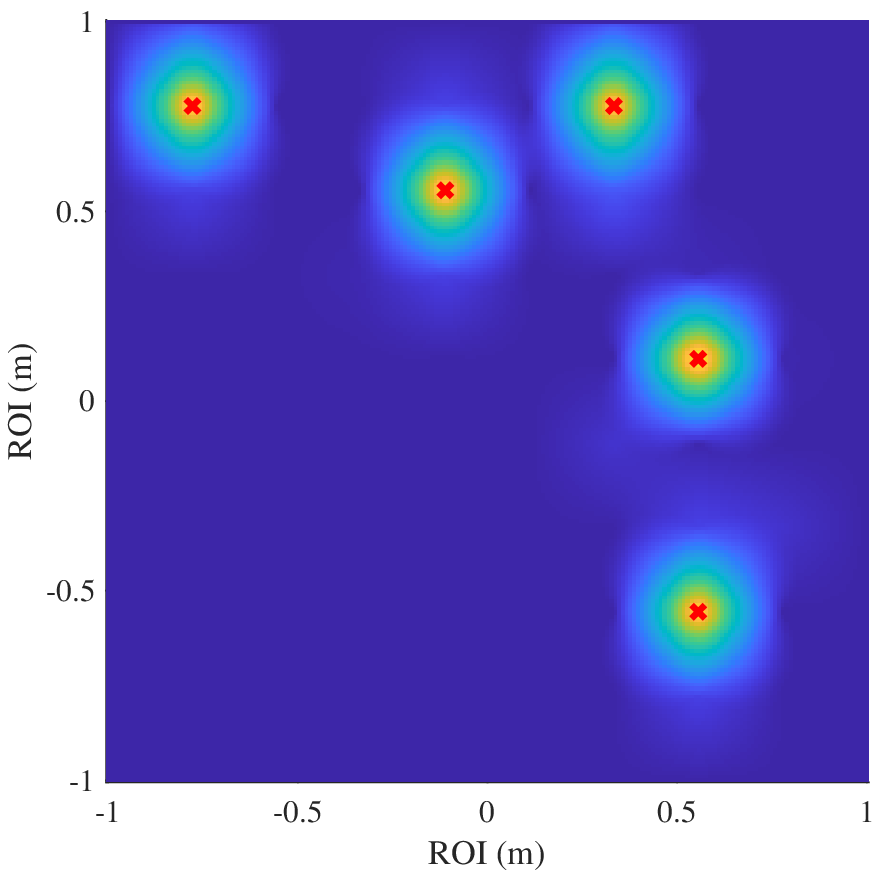}}
  \end{minipage}
  \begin{minipage}[t]{0.22\textwidth}
    \centering
    \subfigure[On-grid imaging with GAMP, PO presented. (${\rm CD} = 0.1120~\rm m$)]{
    \includegraphics[width=0.95\textwidth]{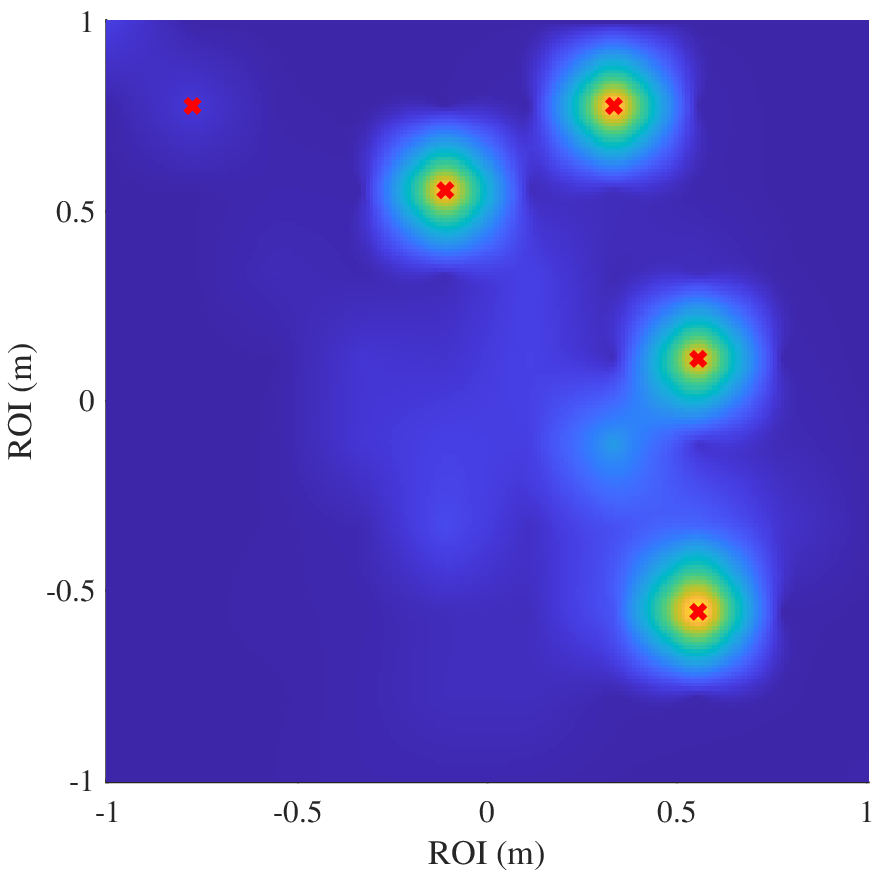}}
    \end{minipage}
  \begin{minipage}[t]{0.22\textwidth}
    \centering
    \subfigure[Off-grid imaging with GAMP, no PO. (${\rm CD} = 0.1612~\rm m$)]{
    \includegraphics[width=0.95\textwidth]{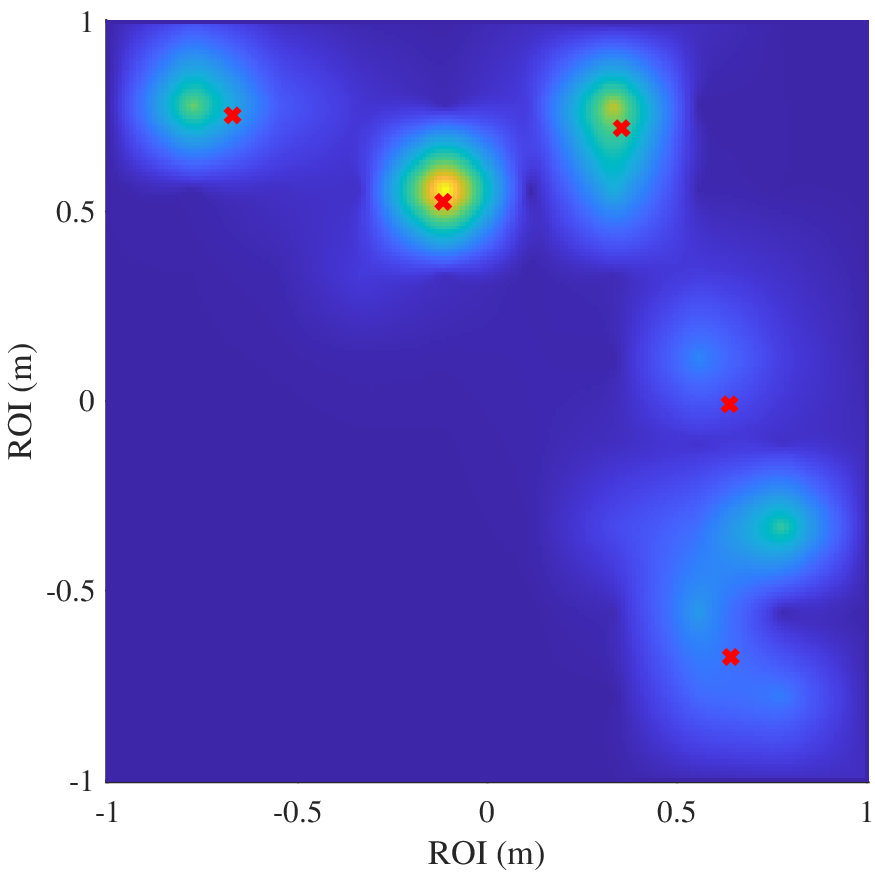}}
    \end{minipage}
    \begin{minipage}[t]{0.22\textwidth}
    \centering
    \subfigure[Off-grid imaging with GAMP, PO presented. (${\rm CD} = 0.1854~\rm m$)]{
    \includegraphics[width=0.95\textwidth]{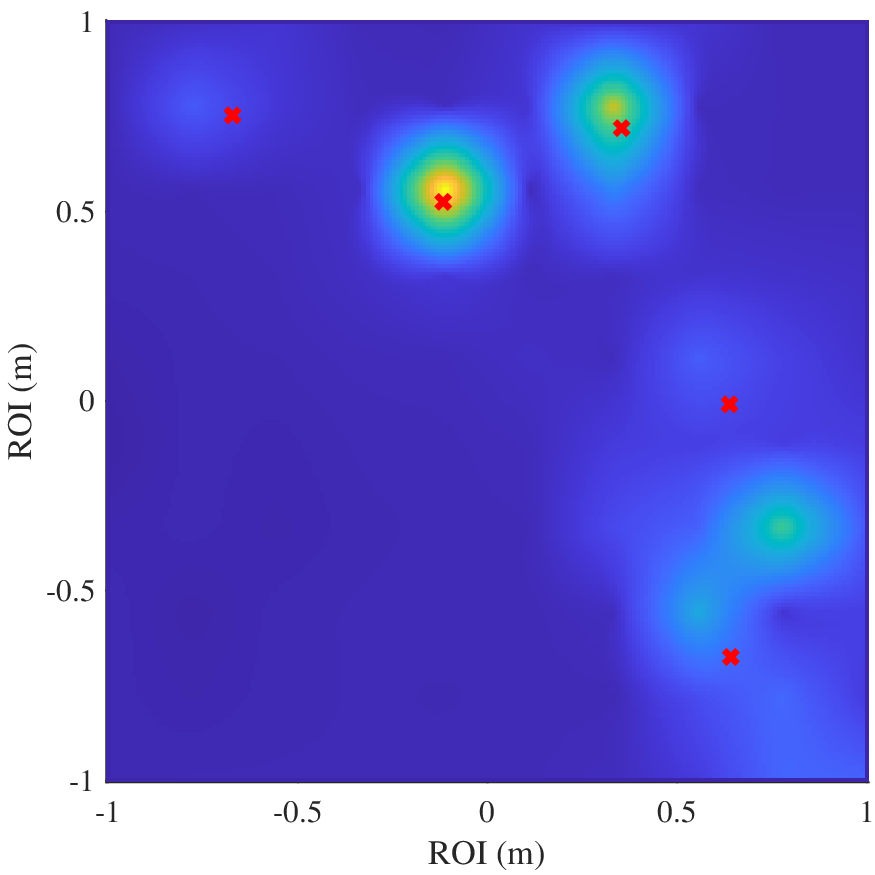}}
    \end{minipage}

  \begin{minipage}[t]{0.22\textwidth}
    \centering
    \subfigure[Off-grid imaging with OG-AMP, no PO, without gradient update. (${\rm CD} = 0.1144~\rm m$)]{
    \includegraphics[width=0.95\textwidth]{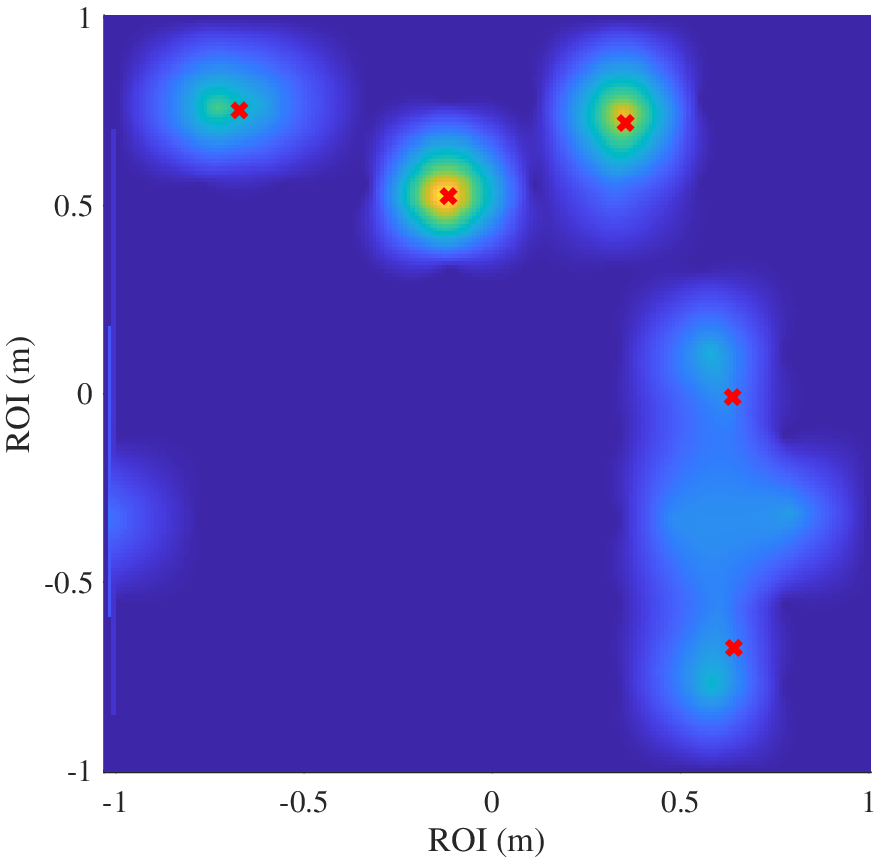}}
    \end{minipage}
  \begin{minipage}[t]{0.22\textwidth}
    \centering
    \subfigure[Off-grid imaging with OG-AMP, no PO, with gradient update. (${\rm CD} = 0.0413~\rm m$)]{
    \includegraphics[width=0.95\textwidth]{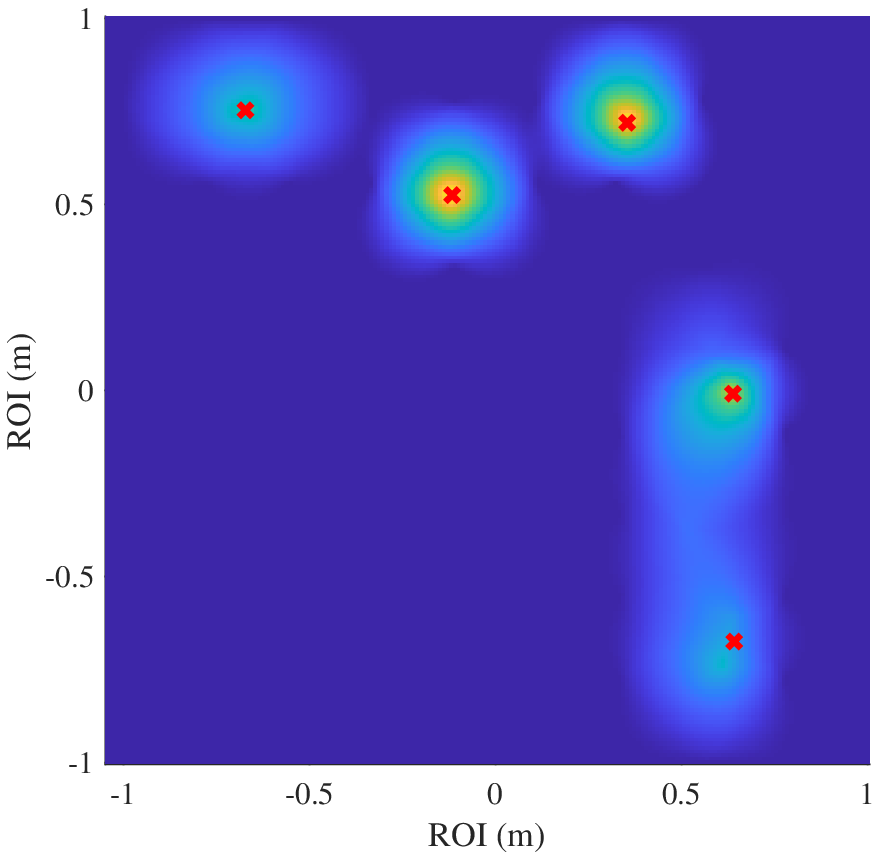}}
    \end{minipage}
  \begin{minipage}[t]{0.22\textwidth}
    \centering
    \subfigure[Off-grid imaging with OG-AMP, PO presented, without AO. (${\rm CD} = 0.1395~\rm m$)]{
    \includegraphics[width=0.93\textwidth]{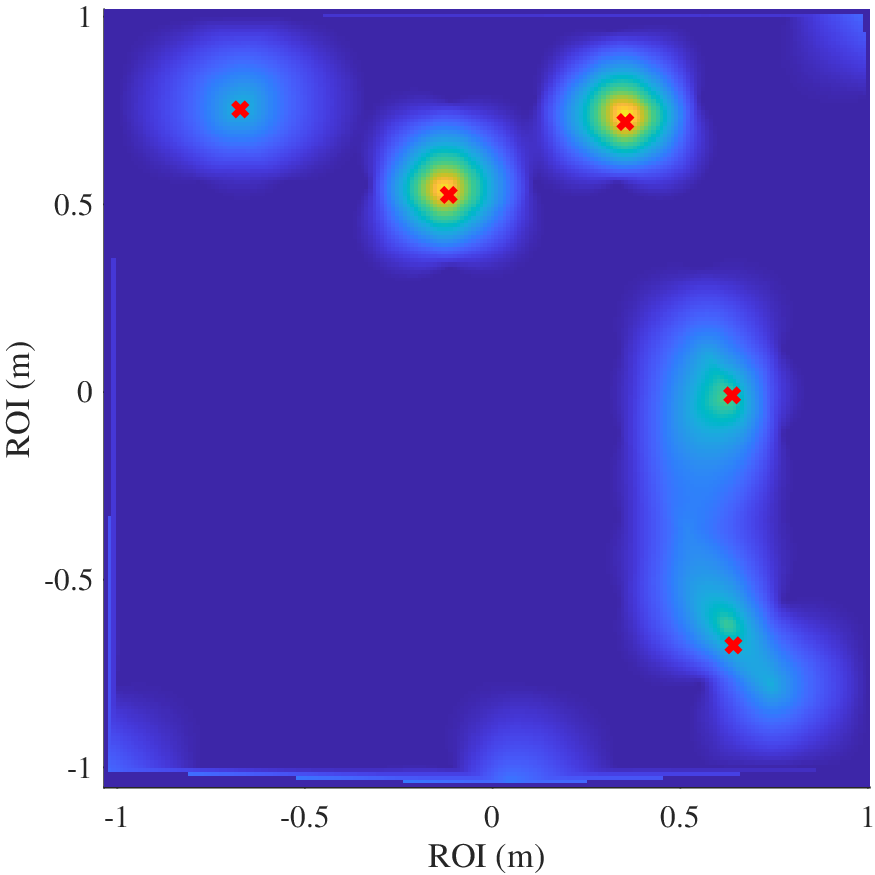}}
    \end{minipage}
  \begin{minipage}[t]{0.22\textwidth}
    \centering
    \subfigure[Off-grid imaging with OG-AMP, PO presented, with AO. (${\rm CD} = 0.0453~\rm m$)]{
    \includegraphics[width=0.93\textwidth]{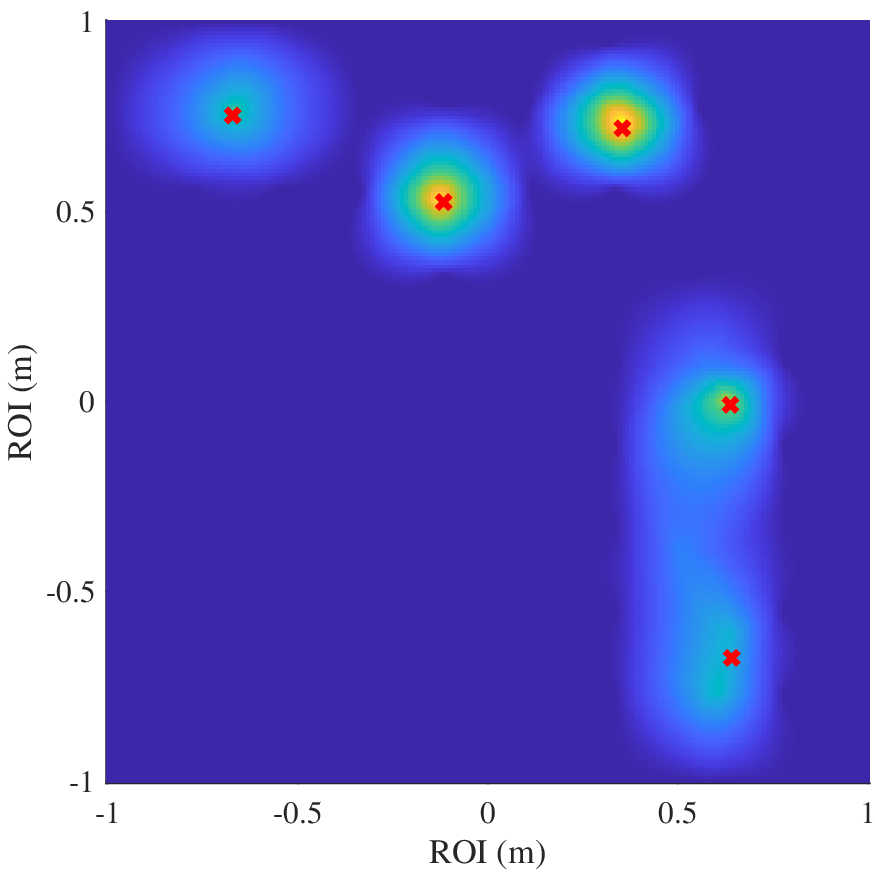}}
    \end{minipage}
\caption{The imaging results of the GAMP method and proposed OG-AMP method. (a)-(d): The performance degradation of GAMP by POs and off-grid offsets. (e)-(f): Effectiveness of OG-AMP iteration and gradient update. (g)-(h): Effectiveness of AO iteration.}
\label{fig:demo}
\vspace{-0.6cm}
\end{figure*}

\section{Numerical Results}
In this section, we evaluate the performance of the proposed NLOS-aided joint OTA synchronization and off-grid imaging framework through numerical simulations. The effectiveness of the proposed method is demonstrated in terms of synchronization accuracy and imaging performance. We also provide a comparison with the representative benchmark scheme GAMP \cite{Rangan}. Compared to OMP, GAMP demonstrates stronger robustness and performance in the compressed sensing reconstruction problem with non-strict Gaussian measurement matrices and it is therefore used as a baseline method for our evaluation \cite{tongxtwc}.

\subsection{Simulation Setting and Metrics}
Unless otherwise specified, all simulation parameters are fixed as follows, except for the variable under study in each performance curve. We consider a distributed MIMO ISAC system with $N_{\rm Tx} = 30$ Txs and $N_{\rm Rx} = 30$ Rxs randomly deployed around the 2D ROI. All nodes are fully connected and have LOS paths. In imaging stage, the ROI is discretized into $N = 25 \times 25$ pixels, with a grid spacing of $0.2 {\rm m}$ in both dimensions. The carrier frequency is set to $f_{\rm c}=1~{\rm GHz}$ and the subcarrier spacing is $\Delta f = 120~{\rm kHz}$.
During the coarse synchronization stage, a full OFDM bandwidth consisting of multiple contiguous subcarriers can be used to enable reliable LOS delay estimation. After coarse synchronization is completed, two representative subcarriers are selected for the subsequent AO-based imaging and fine synchronization process. This is motivated by the fact that imaging resolution in the considered distributed MIMO setup is mainly determined by the effective spatial aperture rather than the bandwidth, so a small number of subcarriers is sufficient while reducing computational complexity.

Each Tx–Rx pair suffers from unknown residual POs after coarse synchronization as described in Section~\ref{sec:coarsesyn}. Based on \eqref{eq:residual_phase_model}, the residual POs of the first subcarrier are generated according to a zero-mean Gaussian distribution with standard deviation $\sigma_\phi = 0.2\pi$. The noise is modeled as complex Gaussian with variance $\sigma^2_{\rm n}$ corresponding to an SNR of $25~{\rm dB}$ unless otherwise specified.
The proposed AO framework iteratively performs off-grid imaging and OTA synchronization. The maximum number of AO iterations is set to $10$, and the error tolerance in the OG-AMP algorithm is set to $\varepsilon = 10^{-4}$. Damping factor $k_{\phi} \in (0,1)$ is set to $0.3$. The prior probability settings for OG-AMP are as follows: $\mu_{x,n} = 1$, $\mu_{s_x,n} = 0$, $\mu_{s_y,n} = 0$, $\sigma_{x,n} = 0.1$. The standard deviations $\sigma_{s_x,n}$, $\sigma_{s_y,n}$ are set to half the pixel size.

Since both the ground truth and the imaging results are off-grid, chamfer distance (CD) is used to measure the spatial accuracy of the imaging results,
{\small\begin{align}
d_{\mathrm{CD}}(\mathcal{X},\mathcal{\hat X}) &= 
\frac{1}{|\mathcal{X}|} \sum_{\mathbf{p}(x)\in \mathcal{X}} \min_{\mathbf{p}(\hat x)\in \mathcal{\hat X}} \|\mathbf{p}(x) - \mathbf{p}(\hat x)\|_2^2 \nonumber
\\ &+ \frac{1}{|\mathcal{\hat X}|} \sum_{\mathbf{p}(\hat x)\in \mathcal{\hat X}} \min_{\mathbf{p}(x)\in \mathcal{X}} \|\mathbf{p}(\hat x) - \mathbf{p}(x)\|_2^2,
\end{align}}\noindent
where $\mathcal{X}$ and $\mathcal{\hat X}$ are pixel sets of the ground truth and imaging results and ${\bf p}(\cdot)$ denotes the spatial coordinate of the pixel.
A smaller $d_{\rm{CD}}$ indicates a closer match between the two pixel sets. Root mean squared error (RMSE) is used to measure the synchronization accuracy, 
% which is defined as
% \begin{equation}
$\mathrm{RMSE} = \frac{1}{MK}\left\|\ln \bm{\Phi} - \ln \hat{\bm{\Phi}} \right\|_{\rm F},$
% \end{equation}
where $\bm \Phi$ and $\hat{\bm \Phi}$ denote the true and estimated POs and $\| \cdot \|_{\rm F}$ represent the Frobenius norm. Considering the periodicity of the phase, unwrapping is required.

\begin{figure*}[t]
  \centering
  \begin{minipage}[t]{0.32\textwidth}
      \centering
      \subfigure[CD of imaging results: After AO, the imaging performance of the PO (red) is close to the perfect calibrated imaging performance (blue).]{
      \includegraphics[width=0.99\textwidth]{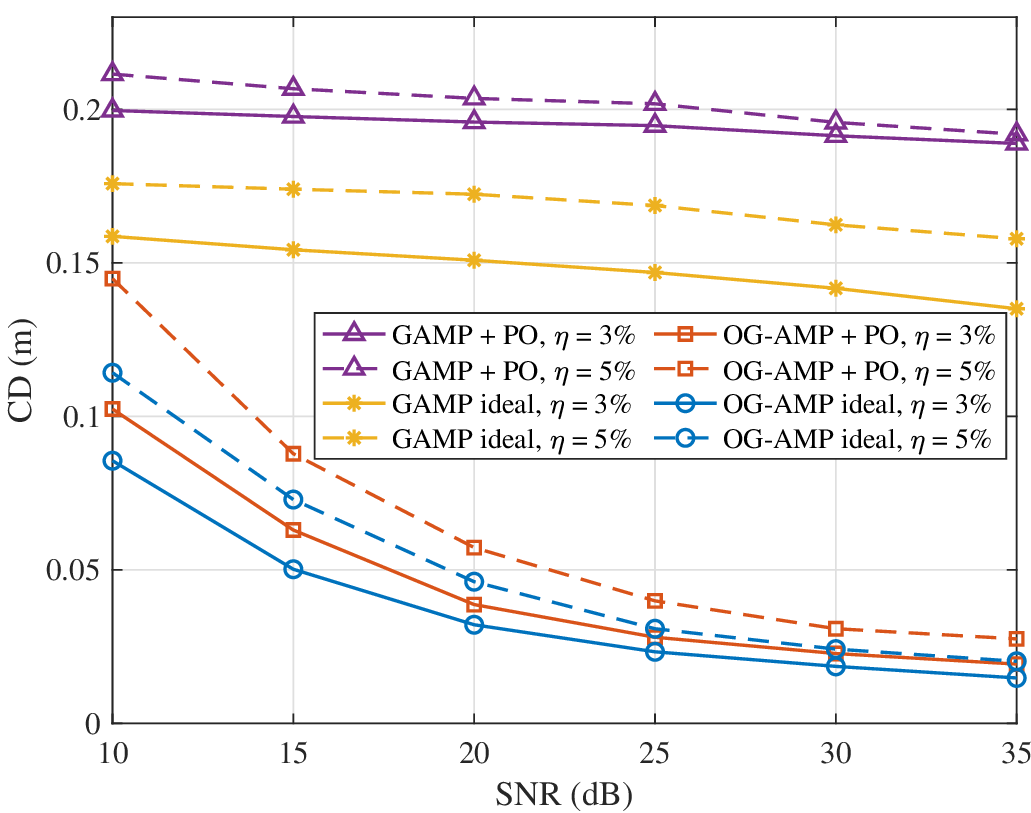}}
  \end{minipage}
  \begin{minipage}[t]{0.02\textwidth}
  ~
    \end{minipage}
  \begin{minipage}[t]{0.32\textwidth}
    \centering
    \subfigure[Estimated RMSE of PO: Comparing the red and blue curves, AO effectively improves synchronization performance.]{
    \includegraphics[width=0.99\textwidth]{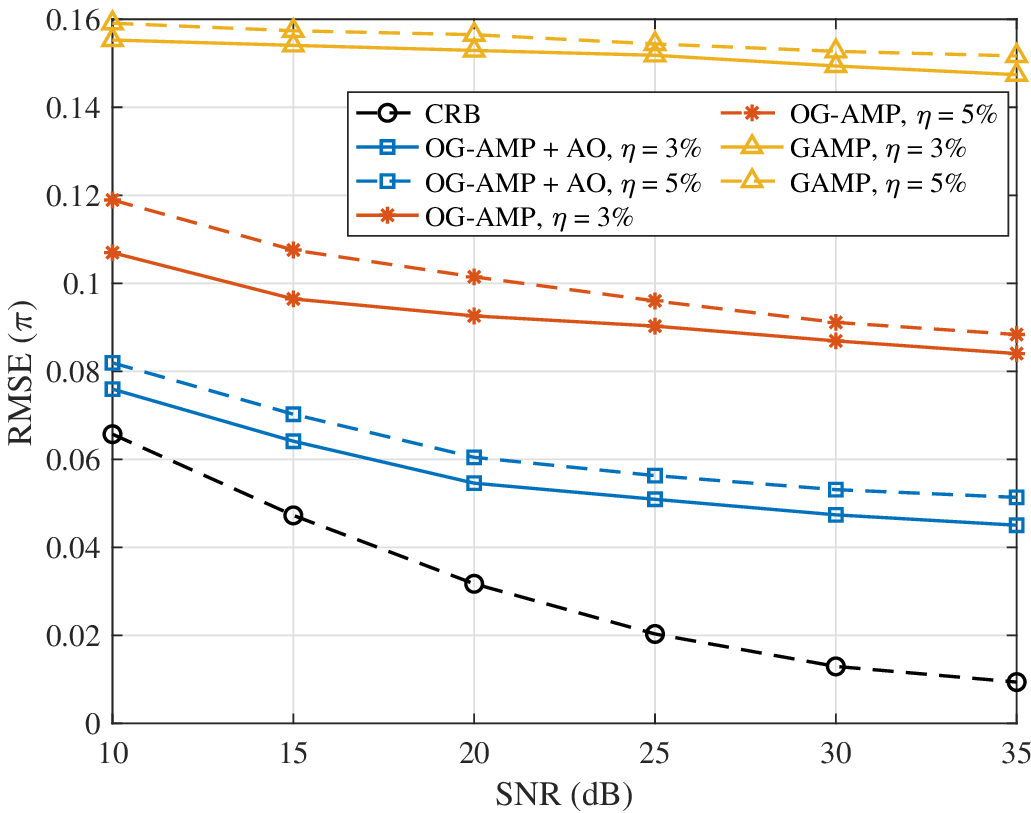}}
\end{minipage}
\caption{The relationship between the SNR and system performance: Overall system performance improves with increasing SNR. CRB curve: Estimating PO based on perfect imaging.}
\label{fig:SNR}
\vspace{-0.6cm}
\end{figure*}

% \begin{figure*}[t]
%   \centering
%   \begin{minipage}[t]{0.4\textwidth}
%       \centering
%       \subfigure[CD of imaging results: After AO, the imaging performance of the PO (red) is close to the perfect calibrated imaging performance (blue).]{
%       \includegraphics[width=0.99\textwidth]{figure/nt_CD.eps}}
%   \end{minipage}
%     \begin{minipage}[t]{0.02\textwidth}
%   ~
%     \end{minipage}
%   \begin{minipage}[t]{0.4\textwidth}
%     \centering
%     \subfigure[Estimated RMSE of PO: Comparing the red and blue curves, AO effectively improves synchronization performance.]{
%     \includegraphics[width=0.99\textwidth]{figure/nt_RMSE.eps}}
% \end{minipage}
% \caption{The relationship between the number of Txs and system performance: Overall system performance improves with increasing $N_{\rm T}$.}
% \label{fig:Tx}
% \vspace{-0.4cm}
% \end{figure*}

\begin{figure*}[t]
  \centering
  \begin{minipage}[t]{0.32\textwidth}
      \centering
      \subfigure[CD of imaging results: The proposed algorithm has better imaging performance.]{
      \includegraphics[width=0.99\textwidth]{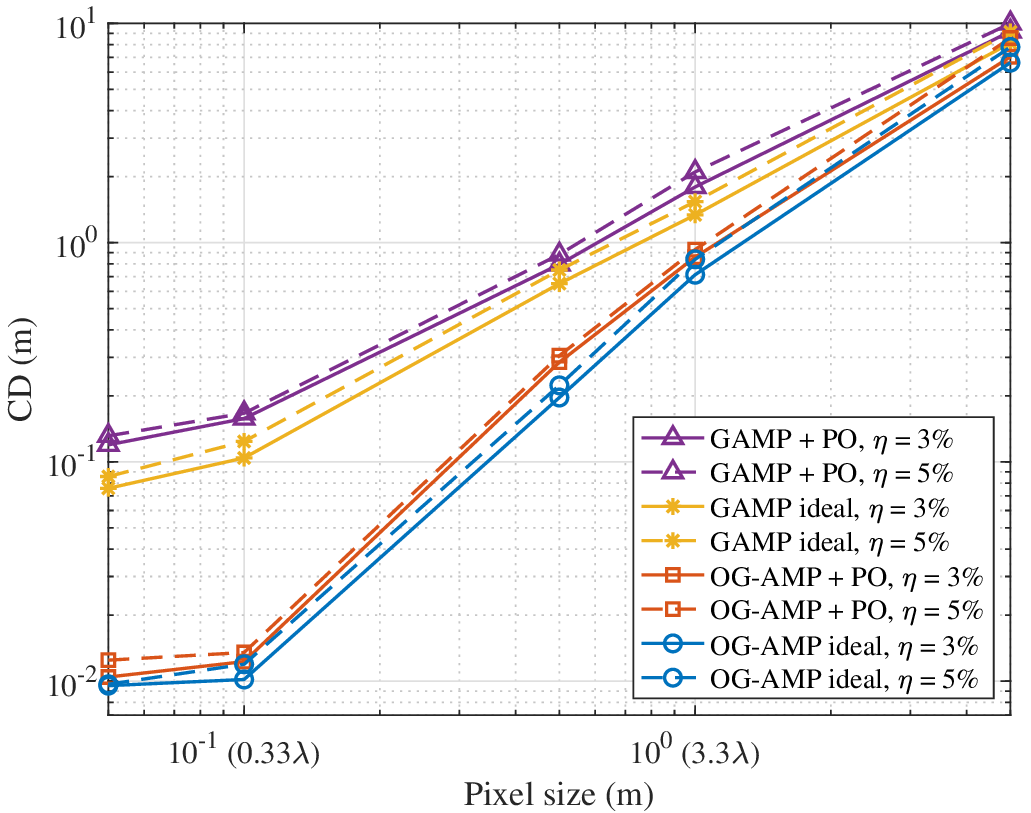}}
  \end{minipage}
      \begin{minipage}[t]{0.02\textwidth}
  ~
    \end{minipage}
  \begin{minipage}[t]{0.32\textwidth}
    \centering
    \subfigure[Estimated RMSE of PO: The proposed AO algorithm supports larger pixel sizes.]{
    \includegraphics[width=0.99\textwidth]{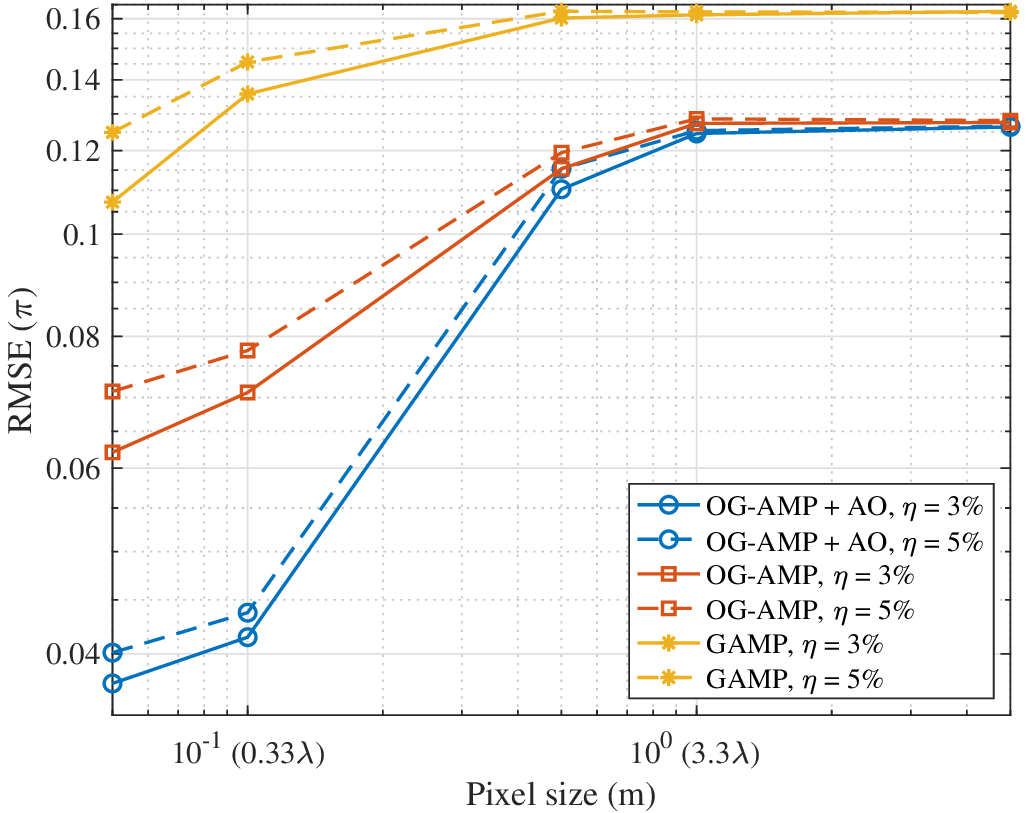}}
\end{minipage}
\caption{The relationship between the pixel size and system performance: }
\label{fig:pixel}
\vspace{-0.4cm}
\end{figure*}

\begin{figure*}[t]
  \centering
  \begin{minipage}[t]{0.32\textwidth}
      \centering
      \subfigure[CD of imaging results: The proposed algorithm exhibits better robustness against PO.]{
      \includegraphics[width=0.99\textwidth]{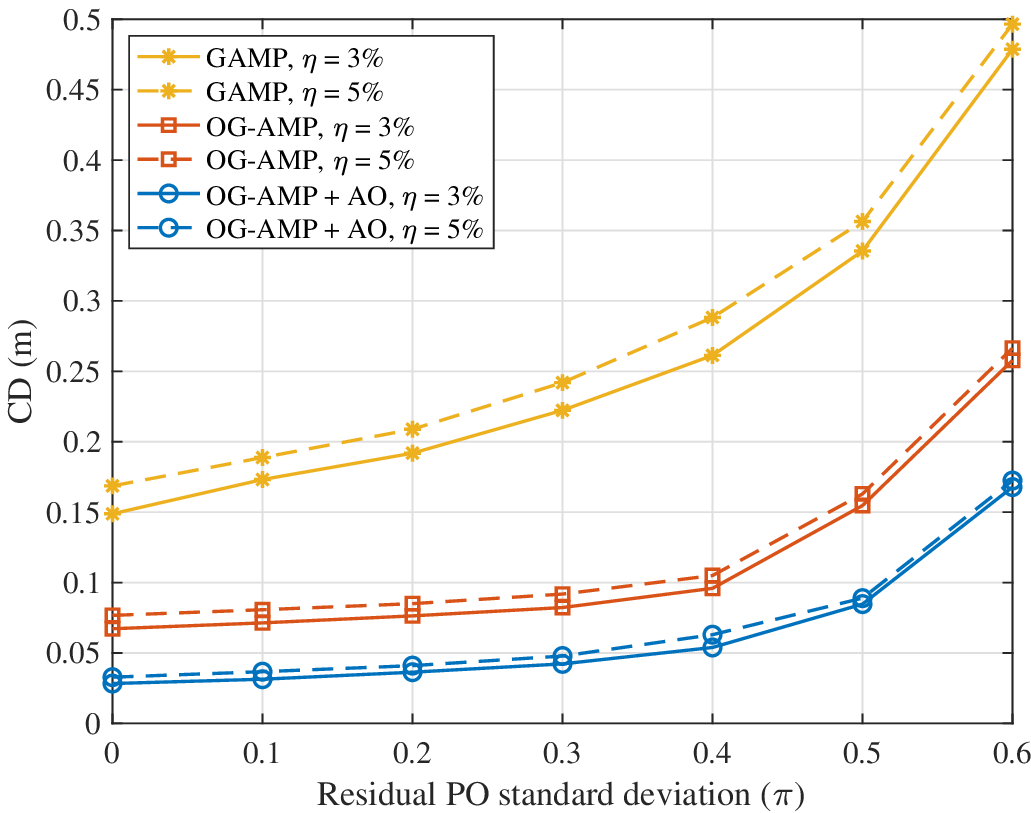}}
  \end{minipage}
      \begin{minipage}[t]{0.02\textwidth}
  ~
    \end{minipage}
  \begin{minipage}[t]{0.32\textwidth}
    \centering
    \subfigure[Estimated RMSE of PO: The conventional algorithm fail quickly, but the proposed algorithm remains robust.]{
    \includegraphics[width=0.99\textwidth]{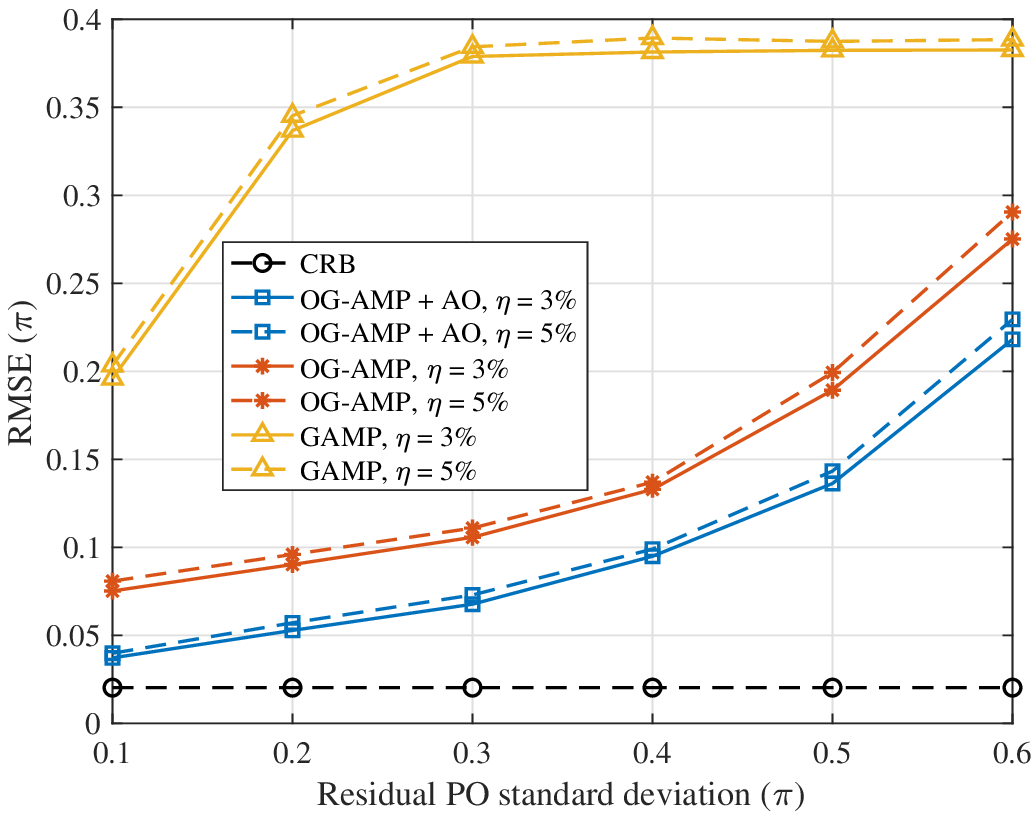}}
\end{minipage}
\caption{The relationship between the residual PO and system performance. $\rm x$-axis: Standard deviation $\sigma_\phi$ of residual PO estimation error after coarse synchronization. $\rm y$-axis: Imaging and refinement synchronization performance in AO.}
\label{fig:PO}
\vspace{-0.4cm}
\end{figure*}

\begin{figure*}[t]
  \centering
  \begin{minipage}[t]{0.32\textwidth}
      \centering
      \subfigure[The convergence performance of OG-AMP.]{
      \includegraphics[width=0.99\textwidth]{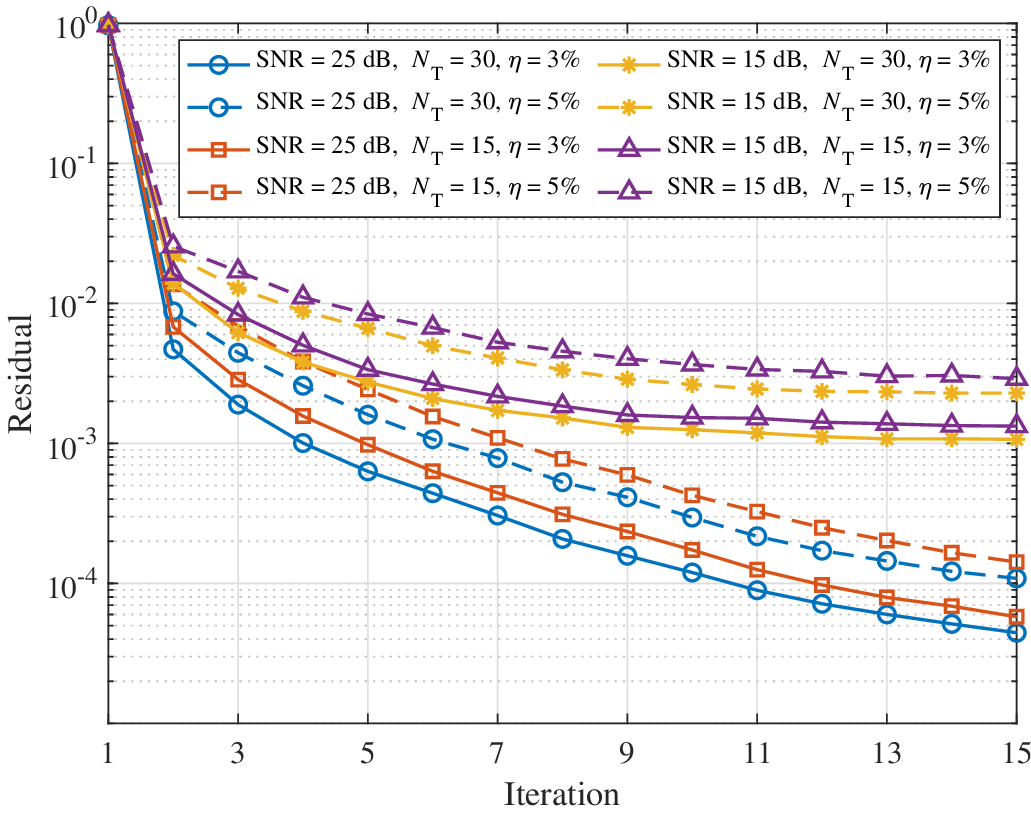}}
  \end{minipage}
  \begin{minipage}[t]{0.32\textwidth}
    \centering
    \subfigure[The convergence performance in CD.]{
    \includegraphics[width=0.99\textwidth]{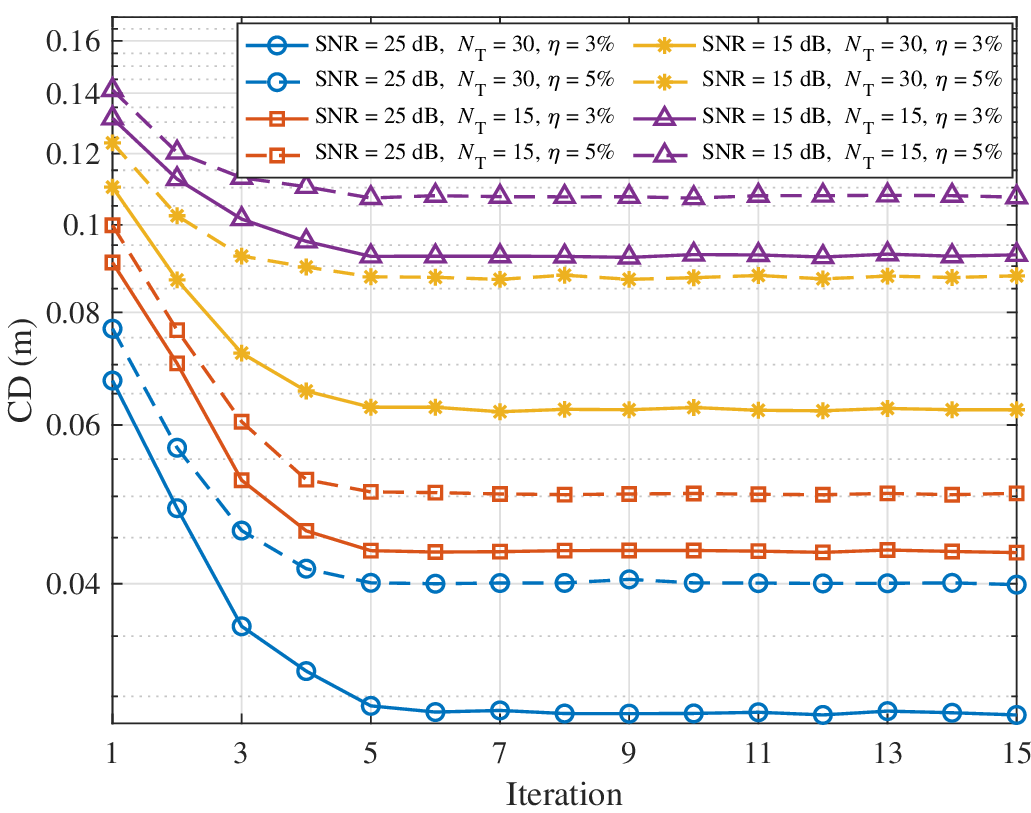}}
\end{minipage}
  \begin{minipage}[t]{0.32\textwidth}
    \centering
    \subfigure[The convergence performance in PO.]{
    \includegraphics[width=0.99\textwidth]{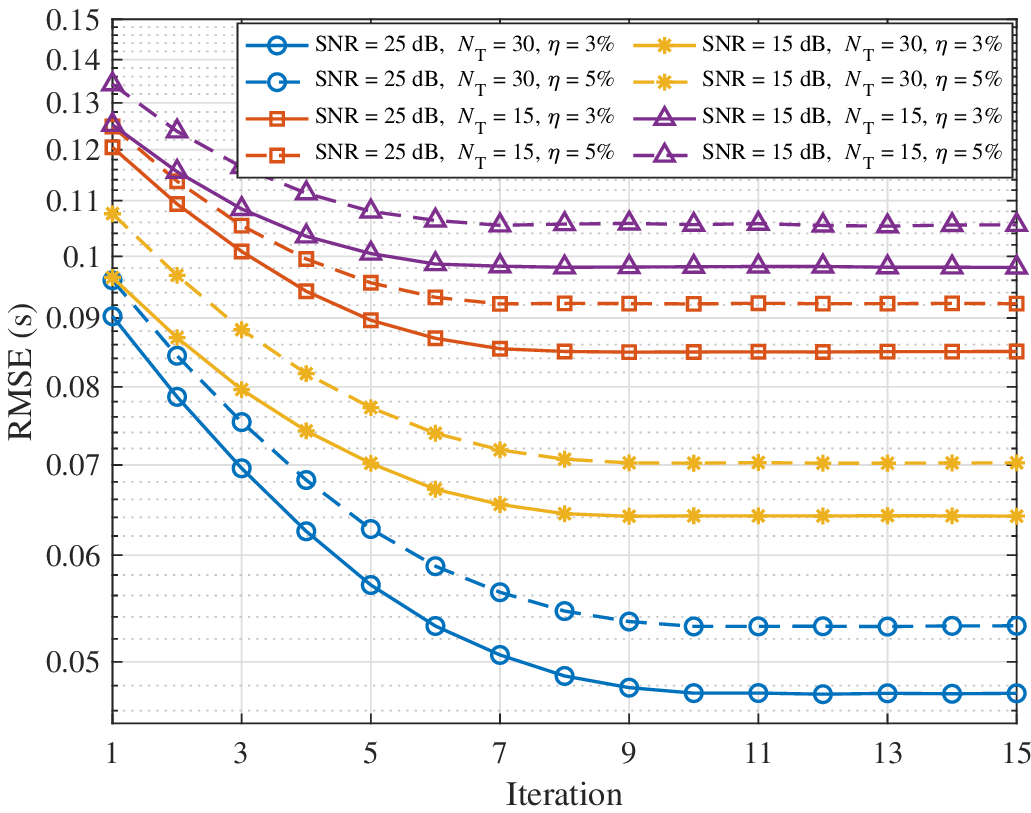}}
\end{minipage}
\caption{The convergence performance of joint synchronization and imaging. (a): The gradient update in the OG-AMP algorithm with $t$ iterations. (b)-(c): Imaging and synchronization in AO with $T$ iterations.}
\label{fig:iter}
\vspace{-0.6cm}
\end{figure*}

\subsection{Performance Evaluation}
Fig.~\ref{fig:demo} shows the intuitive results of both the baseline and the proposed methods. To clearly illustrate the details, we have set up a toy example with $N_{\rm Tx} = 10$ Txs and $N_{\rm Rx} = 10$ Rxs deployed in a $2{\rm m} \times 2{\rm m}$ ROI. The red crosses in the figure represent the location of the ground truth, and the heatmap represents the imaging results. 
Fig.~\ref{fig:demo}(a) through (d) illustrate that while conventional algorithms can achieve good on-grid imaging performance when the target is perfectly centered in the grid, the off-grid effect and the presence of PO greatly degrade the performance. Fig.~\ref{fig:demo}(e) and (f) illustrate the OG-AMP algorithm effectively solves the off-grid imaging problem, based on the gradient update in step 6. Fig.~\ref{fig:demo}(g) and (h) illustrate that the proposed AO algorithm effectively eliminates the influence of PO, making the imaging performance close to the ideal case.

Fig.~\ref{fig:SNR} illustrates the system performance of the baseline and the proposed algorithms under different SNRs. The baseline GAMP algorithm is implemented with the off-grid model proposed in Section~\ref{sec:OGC}; otherwise, it would fail to operate in the presence of off-grid errors. As analyzed in Section~\ref{sec:perB}, the overall system performance improves with increasing SNR. In Fig.~\ref{fig:SNR}(a),  the proposed AO algorithm effectively eliminates the influence of PO, making the imaging performance close to the ideal case.
In Fig.~\ref{fig:SNR}(b), the CRBs are obtained through simulations assuming perfect imaging results. It can be observed that the imaging performance of the baseline GAMP algorithm is insufficient to support NLOS-aided synchronization. In contrast, when PO are present, the proposed AO-based methods consistently outperform the conventional approaches.

% Fig.~\ref{fig:Tx} illustrates the system performance of the baseline and the proposed algorithms under different numbers of Txs. As analyzed in Section~\ref{sec:perB}, the overall system performance improves with an increasing number of Txs due to the improved imaging performance. Similar conclusions to Fig.~\ref{fig:SNR} can be obtained: the proposed algorithm exhibits better performance in both imaging and synchronization.

\subsection{Robustness and Convergence Evaluation}
Fig.~\ref{fig:pixel} illustrates the system performance of the baseline and the proposed algorithms under different pixel sizes. Since the number of pixels is fixed, increasing the pixel size enlarges the ROI. As analyzed in Section~\ref{sec:perB}, larger pixel sizes introduce more off-grid errors, degrading the imaging accuracy and consequently deteriorating the overall system performance. Despite this performance loss, the proposed algorithm consistently outperforms the baseline methods.

Fig.~\ref{fig:PO} illustrates the impact of residual POs after coarse synchronization on the algorithm performance. It can be observed that increasing residual POs degrade the imaging accuracy, which in turn leads to performance loss in synchronization. Nevertheless, compared with conventional computational imaging methods, the proposed approach shows robustness and can operate effectively even with significant residual POs.

Fig.~\ref{fig:iter} illustrates the convergence of the proposed AO algorithm, where the number of iterations is slightly increased for clearer illustration. Fig.~\ref{fig:iter}(a) illustrates that the OG-AMP imaging algorithm also shows fast convergence. However, due to the iteratively gradient updates in step 6, the residual in step 2 continues to decrease slowly. This phenomenon is caused by the gradual movement of pixel center points toward the true target locations. Nevertheless, this refinement does not lead to further improvement in the overall system performance. Therefore, an appropriate stopping threshold $\varepsilon$ should be selected to avoid unnecessary iterations. As shown in Figs.~\ref{fig:iter}(b) and \ref{fig:iter}(c), the system performance converges rapidly within a few AO iterations.

\section{Conclusion}
This paper proposed an NLOS-aided joint framework for OTA phase synchronization and off-grid imaging in distributed MIMO ISAC systems. Based on LOS-assisted coarse synchronization, a unified ISAC framework is proposed by combining off-grid computational imaging with NLOS-assisted refined synchronization within a closed-loop AO framework.
Analytical results provided insights into the impact of carrier frequency, bandwidth, and grid resolution on off-grid modeling errors. Numerical simulations demonstrated that the proposed approach achieves accurate synchronization and imaging under POs. These results highlight the robustness and practical applicability of the proposed method for large-scale distributed ISAC systems.

\begin{appendices}
\section{Approximation of Message $\mu_{x_{n}\rightarrow {\rm F}_{\hat y,m}}^{(t+1)}(x_{n})$}\label{ap:A}
According to Fig.~\ref{fig:fg}, the message $\mu_{x_{n}\rightarrow {\rm F}_{\hat y,m}}^{(t+1)}(x_{n})$ can be expressed as 
% \begin{align}
    $\mu^{(t)}_{{\rm F}_{\hat y,m} \rightarrow x_n }(x_n) \propto 
    \int \Pr(\hat y_m|\tilde{\bm x}, \tilde{\bf A}) $ 
    $\prod_{i \neq n}\mu^{(t)}_{x_i \rightarrow {\rm F}_{\hat y,m}}(x_i){\rm d}x_i 
    \prod_{l = 1}^N\mu^{(t)}_{s_{x,l} \rightarrow {\rm F}_{\hat y,m}}(s_{x,l}){\rm d}s_{x,l} $ 
    $\prod_{l = 1}^N\mu^{(t)}_{s_{y,l} \rightarrow {\rm F}_{\hat y,m}}(s_{y,l}){\rm d}s_{y,l}.$
% \end{align}
We isolate the $x_n$ term,
% \begin{align}
    $\hat y_m = \; a_{m,n} x_n + \Big( \sum_{j \ne n} a_{m,j} x_j + \sum_{l=1}^N a^{(t)}_{x,m,l} s_{x,l} 
    + \sum_{l=1}^N a^{(t)}_{y,m,l} s_{y,l} + n_m \Big) 
     = \; a_{m,n} x_n + S^{(t)}_{m \rightarrow x_n },$
% \end{align}
where $S^{(t)}_{m \rightarrow x_n }$ is a sum of many random variables, the distribution of each variable $j$ is given by its incoming message $\mu_{j \rightarrow {\rm F}_{\hat y,m}}^{(t)}$. Following the standard assumption ($M,N\rightarrow\infty, M/N = {\rm const}$) in belief propagation for such models, we apply the central limit theorem. We approximate this entire sum as a complex Gaussian random variable. The message is simplified to
% \begin{align}
$\mu_{{\rm F}_{\hat y,m}\rightarrow x_{n}}^{(t)}(x_n) \propto \int \mathcal{CN}(\hat y_m ; a_{m,n} x_n + S_{m \rightarrow x_n}^{(t)}, \sigma_{\rm n}^2)
 \Pr(S_{{m} \rightarrow x_n}^{(t)}) {\rm d}S_{{m} \rightarrow x_n}^{(t)} 
\approx \int \mathcal{CN}(\hat y_m ; a_{m,n} x_n + S_{{m} \rightarrow x_n}^{(t)}, \sigma_{\rm n}^2) \mathcal{CN}(S_{{m} \rightarrow x_n}^{(t)} ; Z_{{m} \rightarrow x_n}^{(t)}, V_{{m} \rightarrow x_n}^{(t)}) {\rm d}S_{{m} \rightarrow x_n}^{(t)}
\propto \;  \mathcal{CN}(x_n; \hat x_{{m} \rightarrow x_n}^{(t)}, v_{{m} \rightarrow x_n}^{(t)}),$
% \end{align}
where $\hat x_{{m} \rightarrow x_n}^{(t)}$ and $v_{{m} \rightarrow x_n}^{(t)}$ are the mean and variance of $\mu_{{\rm F}_{\hat y,m}\rightarrow x_{n}}^{(t)}(x_n)$ as calculated in \eqref{eq:xmtxn} and \eqref{eq:vmtxn}. The mean $Z_{{m} \rightarrow x_n}^{(t)}$ and variance $V_{{m} \rightarrow x_n}^{(t)}$ of $S_{{m} \rightarrow x_n}^{(t)}$ are calculated as
{\small\begin{align}
Z_{m\rightarrow x_{n}}^{(t)} & =\sum_{j\ne n}a_{m,j} \hat{x}_{j \to {\rm F}_{\hat y,m}}^{(t)} + \sum_{l=1}^{N}a^{(t)}_{x,m,l} \hat{s}_{x,l \to {\rm F}_{\hat y,m}}^{(t)} \nonumber\\ &+ \sum_{l=1}^{N}a^{(t)}_{y,m,l} \hat{s}_{y,l \to {\rm F}_{\hat y,m}}^{(t)},
\label{eq:zmtxn}
\end{align}}\noindent
{\small\begin{align}
V_{m\rightarrow x_{n}}^{(t)} &= \sum_{j\ne n}|a_{m,j}|^2 v_{j \to {\rm F}_{\hat y,m}}^{(t)} + \sum_{l=1}^{N}|a^{(t)}_{x,m,l}|^2 v_{x,l \to {\rm F}_{\hat y,m}}^{(t)} \nonumber\\ &+ \sum_{l=1}^{N}|a^{(t)}_{y,m,l}|^2 v_{y,l \to {\rm F}_{\hat y,m}}^{(t)},
\label{eq:vvmtxn}
\end{align}}\noindent
where $\hat{x}_{j \to {\rm F}_{\hat y,m}}^{(t)}$ and $v_{j \to {\rm F}_{\hat y,m}}^{(t)}$ are the mean and variance of the message $\mu_{x_j \rightarrow {\rm F}_{\hat y,m}}^{(t)}(x_j)$, which will be derived later.

\section{Approximation of Message $\mu_{{\rm F}_{x,n}\rightarrow x_{n}}^{(t+1)}(x_{n})$}\label{ap:B}
The message $\mu_{{x_n} \rightarrow {\rm F}_{x,n}}^{(t)}(x_n)$ in $t$-th iteration is approximated as
% \begin{align}
    $ \mu_{{x_n} \rightarrow {\rm F}_{x,n}}^{(t)}(x_n) \propto \prod_{m=1}^{M}\mu_{{\rm F}_{\hat y,m} \rightarrow x_n}^{(t)}(x_n) = \prod_{m=1}^{M} \mathcal{CN}(x_n; \hat x_{{m} \rightarrow x_n}^{(t)}, v_{{m} \rightarrow x_n}^{(t)}) \propto \mathcal{CN}(x_n; \hat{x}_{n \to {\rm F}_x}^{(t)}, v_{n \to {\rm F}_x}^{(t)}),$
% \end{align}
where $\hat{x}_{n \to {\rm F}_x}^{(t)}$ and $v_{n \to {\rm F}_x}^{(t)}$ are the mean and variance of $\mu_{x_{n} \rightarrow {\rm F}_{x,n}}^{(t)}(x_n)$, then we obtain \eqref{eq:xntfx} and \eqref{eq:vntfx}.

\section{Derivation of $Z_{x,n}^{(t)}(c_n)$}\label{ap:C}
When $c_n = 1$: $\Pr(x_n|c_n=1) = \mathcal{CN}(x_n; \mu_{x,n}, \sigma_{x,n}^2)$ and $\mu_{x_{n}\rightarrow {\rm F}_{x,n}}^{(t)}(x_n) \propto \mathcal{CN}(x_n; \hat{x}_{n\rightarrow {\rm F}_x}^{(t)}, v_{n\rightarrow {\rm F}_x}^{(t)})$, we obtain
    % \begin{align}
        $Z_{x,n}^{(t)}(1) = \int \mathcal{CN}(x_n; \mu_{x,n}, \sigma_{x,n}^2) \cdot \mathcal{CN}(x_n; \hat{x}_{n\rightarrow {\rm F}_x}^{(t)}, v_{n\rightarrow {\rm F}_x}^{(t)}) {\rm d}x_n
        = \mathcal{CN}(\mu_{x,n} ; \hat{x}_{n\rightarrow {\rm F}_x}^{(t)}, \sigma_{x,n}^2 + v_{n\rightarrow {\rm F}_x}^{(t)}).$    % \end{align}
        
When $c_n = 0$: $\Pr(x_n|c_n=0) = \delta(x_n)$, we obtain
    % \begin{align}
        $Z_{x,n}^{(t)}(0) = \int \delta(x_n)\mathcal{CN}(x_n; \hat{x}_{n\rightarrow {\rm F}_x}^{(t)}, v_{n\rightarrow {\rm F}_x}^{(t)}) {\rm d}x_n
        = \mathcal{CN}(0 ; \hat{x}_{n\rightarrow {\rm F}_x}^{(t)}, v_{n\rightarrow {\rm F}_x}^{(t)}).$
    % \end{align}

\section{Approximation of Message $\mu_{{\rm F}_{x,n}\rightarrow x_{n}}^{(t+1)}(x_{n})$}\label{ap:D}
The message $\mu_{{\rm F}_{x,n}\rightarrow x_{n}}^{(t+1)}(x_{n})$ in $(t+1)$-th iteration is expressed as
{\small\begin{align}
    & \mu^{(t+1)}_{F_{x,n} \rightarrow x_{n}}(x_n) \nonumber\\= \; & \sum_{c_n \in \{0,1\}} \Pr(x_n | c_n)\mu_{c_n \to {\rm F}_{x,n}}(c_n)  Z_{s_{x,n}}^{(t)}(c_{n}) Z_{s_{y,n}}^{(t)}(c_{n}),
    \label{eq:mfxntxn}
\end{align}}\noindent
where $Z_{s_{x,n}}^{(t)}(c_{n})$ and $Z_{s_{y,n}}^{(t)}(c_{n})$ are defined in \eqref{eq:mfymtcn}. The derivations of ${s}_{x,n}$ and ${s}_{y,n}$ are similar. 

When $c_n = 1$: $\Pr(x_n | c_n=1) = \mathcal{CN}(x_{n};\mu_{x,n},\sigma_{x,n}^{2})$ and $\mu_{c_n \to {\rm F}_{x,n}}(1) \propto \eta$, we obtain the term in \eqref{eq:mfxntxn},
% \begin{equation}
    ${\rm Term}_1 = \mathcal{CN}(x_{n};\mu_{x,n},\sigma_{x,n}^{2}) \cdot \eta \cdot Z_{s_{x,n}}^{(t)}(1) \cdot Z_{s_{y,n}}^{(t)}(1).$
% \end{equation}
    
When $c_n = 0$: $\Pr(x_n | c_n=0) = \delta(x_n)$ and $\mu_{c_n \to {\rm F}_{x,n}}(0) \propto (1-\eta)$, we obtain the term in \eqref{eq:mfxntxn}, 
% \begin{equation}
    ${\rm Term}_0 = \delta(x_n) \cdot (1-\eta) \cdot Z_{s_{x,n}}^{(t)}(0) \cdot Z_{s_{y,n}}^{(t)}(0).$
% \end{equation}
Define weights as \eqref{eq:kn0} and \eqref{eq:kn1}, the passed message from FN ${\rm F}_x$ to VN $\tilde{\bm x}$ is calculated as the sum of above terms, then we obtain \eqref{eq:fxntxn1}.

\section{Approximation of Message $\mu_{x_n \rightarrow {\rm F}_{\hat y,m}}^{(t+1)}(x_n)$}\label{ap:E}
The message $\mu_{x_n \rightarrow {\rm F}_{\hat y,m}}^{(t+1)}(x_n)$ in $(t+1)$-th iteration is approximated as
\begin{align}
    \mu_{x_{n}\rightarrow {\rm F}_{\hat y,m}}^{(t+1)}(x_{n}) & \propto \mu_{{\rm F}_{x,n}\rightarrow x_{n}}^{(t+1)}(x_{n}) \cdot \prod_{k \ne m} \mu_{{\rm F}_{\hat y,k} \to x_n}^{(t)}(x_n),
    \label{eq:fyktoxn}
\end{align}
where
{\small \begin{align}
    \prod_{k \ne m} \mu_{{\rm F}_{\hat y,k} \to x_n}^{(t)}(x_n) &\propto \frac{\mu_{x_{n}\rightarrow {\rm F}_{x,n}}^{(t)}(x_{n})}{\mu_{{\rm F}_{\hat y,m}\rightarrow x_{n}}^{(t)}(x_{n})} \\
    &\propto \frac{\mathcal{CN}(x_{n};\hat{x}_{n\rightarrow {\rm F}_{x}}^{(t)},v_{n\rightarrow {\rm F}_{x}}^{(t)})}{\mathcal{CN}(x_{n};\hat{x}_{m\rightarrow x_{n}}^{(t)},v_{m\rightarrow x_{n}}^{(t)})}\\
    &\propto \mathcal{CN}(x_n; \hat{x}_{{\rm cav}, m}^{(t)}, v_{{\rm cav}, m}^{(t)}).
    \label{eq:fyktoxnem}
\end{align}}\noindent
The mean $\hat{x}_{{\rm cav}, m}^{(t)}$ and $v_{{\rm cav}, m}^{(t)}$ variance are calculated in \eqref{eq:xcav} and \eqref{eq:vcav}. Plug \eqref{eq:fxntxn1} and \eqref{eq:fyktoxnem} into \eqref{eq:fyktoxn} , we obtain
% \begin{align}
$ \mu_{x_{n}\rightarrow {\rm F}_{\hat y,m}}^{(t+1)}(x_{n})  
    \propto \Big[ k_{n,0}^{(t+1)} \delta(x_n) + k_{n,1}^{(t+1)} \mathcal{CN}(x_{n};\mu_{x,n},\sigma_{x,n}^{2}) \Big] \cdot  \mathcal{CN}(x_n; \hat{x}_{{\rm cav}, m}^{(t)}, v_{{\rm cav}, m}^{(t)})
    \propto   W_{n,0}^{(t+1)} \cdot \delta(x_n) + W_{n,1}^{(t+1)} \cdot \mathcal{CN}(x_n; \hat{x}_{{\rm prod}, m}^{(t+1)}, v_{{\rm prod}, m}^{(t+1)}),$
% \end{align}
where weights $ W_{n,0}^{(t+1)}$ and $W_{n,1}^{(t+1)}$ are calculated in \eqref{eq:wn0} and \eqref{eq:wn1}. The mean $\hat{x}_{{\rm prod}, m}^{(t+1)}$ and variance $v_{{\rm prod}, m}^{(t+1)}$ are calculated in \eqref{eq:xprod} and \eqref{eq:vprod}. Based on the normalized weights $P_{n,1 \to m}^{(t+1)}$ defined in \eqref{eq:pn1}, the mean and variance of message $\mu_{x_n \rightarrow {\rm F}_{\hat y,m}}^{(t+1)}(x_n)$ are derived as
% \begin{align}
    $\hat{x}_{n \to {\rm F}_{\hat y,m}}^{(t+1)} = P_{n,0 \to m}^{(t+1)} \cdot {\rm E}[\delta(x_n)] + P_{n,1 \to m}^{(t+1)} \cdot {\rm E}[\mathcal{CN}(x_n; \hat{x}_{{\rm prod}, m}^{(t+1)}, v_{{\rm prod}, m}^{(t+1)})] = P_{n,1 \to m}^{(t+1)} \cdot \hat{x}_{{\rm prod}, m}^{(t+1)},$
% \end{align}
% \begin{align}
    $v_{n \to {\rm F}_{\hat y,m}}^{(t+1)} = {\rm E}[|x_n|^2] -  |{\rm E}[x_n]|^2 = \Big[ P_{n,1 \to m}^{(t+1)} (v_{{\rm prod}, m}^{(t+1)} + |\hat{x}_{{\rm prod}, m}^{(t+1)}|^2) \Big] - \Big[ (P_{n,1 \to m}^{(t+1)})^2 |\hat{x}_{{\rm prod}, m}^{(t+1)}|^2 \Big] = P_{n,1 \to m}^{(t+1)} v_{{\rm prod}, m}^{(t+1)} + \left( P_{n,1 \to m}^{(t+1)} - (P_{n,1 \to m}^{(t+1)})^2 \right) |\hat{x}_{{\rm prod}, m}^{(t+1)}|^2.$
% \end{align}
where the normalized weights $P_{n,1 \to m}^{(t+1)}$ is calculated as
\begin{equation}\small
    P_{n,1 \to m}^{(t+1)} = \frac{W_{n,1}^{(t+1)}}{W_{n,0}^{(t+1)} + W_{n,1}^{(t+1)}}.
    \label{eq:pn1}
\end{equation}
Weights $ W_{n,0}^{(t+1)}$ and $W_{n,1}^{(t+1)}$ are defined as 
\begin{equation}\small
    W_{n,0}^{(t+1)} \propto k_{n,0}^{(t+1)} \cdot \mathcal{CN}(0; \hat{x}_{{\rm cav}, m}^{(t)}, v_{{\rm cav}, m}^{(t)}),
    \label{eq:wn0}
\end{equation}
\begin{equation}\small
    W_{n,1}^{(t+1)} \propto k_{n,1}^{(t+1)} \cdot \mathcal{CN}(\mu_{x,n}; \hat{x}_{{\rm cav}, m}^{(t)}, \sigma_{x,n}^2 + v_{{\rm cav}, m}^{(t)}).
    \label{eq:wn1}
\end{equation}
Means $\hat{x}_{{\rm prod}, m}^{(t+1)}$, $\hat{x}_{{\rm cav}, m}^{(t)}$, and variances $v_{{\rm prod}, m}^{(t+1)}$, $v_{{\rm cav}, m}^{(t)}$ are defined as
\begin{equation}\small
    \hat{x}_{{\rm prod}, m}^{(t+1)} = v_{{\rm prod}, m}^{(t+1)} \Big( \frac{\mu_{x,n}}{\sigma_{x,n}^2} + \frac{\hat{x}_{{\rm cav}, m}^{(t)}}{v_{{\rm cav}, m}^{(t)}} \Big),
    \label{eq:xprod}
\end{equation}
\begin{equation}\small
    v_{{\rm prod}, m}^{(t+1)} = \Big(\frac{1}{\sigma_{x,n}^2} + \frac{1}{v_{{\rm cav}, m}^{(t)}}\Big)^{-1},
    \label{eq:vprod}
\end{equation}
\begin{equation}\small
    \hat{x}_{{\rm cav}, m}^{(t)} = v_{{\rm cav}, m}^{(t)} \Big( \frac{\hat{x}_{n\rightarrow {\rm F}_{x}}^{(t)}}{v_{n\rightarrow {\rm F}_{x}}^{(t)}} - \frac{\hat{x}_{m\rightarrow x_{n}}^{(t)}}{v_{m\rightarrow x_{n}}^{(t)}} \Big),
    \label{eq:xcav}
\end{equation}
\begin{equation}\small
    v_{{\rm cav}, m}^{(t)} = \Big(\frac{1}{v_{n\rightarrow F_{x}}^{(t)}} - \frac{1}{v_{m\rightarrow x_{n}}^{(t)}}\Big)^{-1}.
    \label{eq:vcav}
\end{equation}

\end{appendices}

\ifCLASSOPTIONcaptionsoff
  \newpage
\fi

\bibliographystyle{IEEEbib}
\bibliography{IEEEabrv, ref}
\end{document}